\documentclass[12pt,twoside, a4paper]{article}
\def\pd{\partial}
\def\mc{\mathcal}

\usepackage[dvips]{graphicx}
\usepackage{amssymb}
\usepackage{amssymb,amsmath}
\usepackage{graphicx}
\usepackage{dsfont}
\usepackage{caption}
\usepackage{subcaption}
\usepackage{mathtools}
\usepackage{verbatim}
\usepackage{graphicx}
\usepackage{multirow}
\usepackage[outline]{contour}
\usepackage{xcolor,colortbl}
\input{epsf.sty}  
\begin{document}
\begin{center}
\Large{\textbf{$AdS_5$ vacua and holographic RG flows from 5D $N=4$ gauged supergravity}}
\end{center}
\vspace{1 cm}
\begin{center}
\large{\textbf{Parinya Karndumri}}
\end{center}
\begin{center}
String Theory and Supergravity Group, Department
of Physics, Faculty of Science, Chulalongkorn University, 254 Phayathai Road, Pathumwan, Bangkok 10330, Thailand \\
E-mail: parinya.ka@hotmail.com \vspace{1 cm}
\end{center}
\begin{abstract}
We study five-dimensional $N=4$ gauged supergravity coupled to five vector multiplets with $SO(2)_D\times SO(3)\times SO(3)$ gauge group. There are four supersymmetric $AdS_5$ vacua in the truncation to $SO(2)_{\textrm{diag}}$ invariant scalars. Two of these vacua preserve the full $N=4$ supersymmetry with $SO(2)_D\times SO(3)\times SO(3)$ and $SO(2)_D\times SO(3)_{\textrm{diag}}$ symmetries. These have an analogue in $N=4$ gauged supergravity with $SO(2)\times SO(3)\times SO(3)$ gauge group. The other two $AdS_5$ vacua preserve only $N=2$ supersymmetry with $SO(2)_{\textrm{diag}}\times SO(3)$ and $SO(2)_{\textrm{diag}}$ symmetry. The former has an analogue in the previous study of $SO(2)_D\times SO(3)$ gauge group while the latter is a genuinely new $N=2$ $AdS_5$ vacuum. These vacua should be dual to $N=2$ and $N=1$ superconformal field theories (SCFTs) in four dimensions with different flavour symmetries. We give the full scalar mass spectra at all of the $AdS_5$ critical points which provide information on conformal dimensions of the dual operators. Finally, we study holographic RG flows interpolating between these $AdS_5$ vacua and find a new class of solutions. In addition to the RG flows from the trivial $SO(2)_D\times SO(3)\times SO(3)$ $N=4$ critical point, at the origin of the scalar manifold, to all the other critical points, there is a family of RG flows from the trivial $N=4$ critical point to the new $SO(2)_{\textrm{diag}}$ $N=2$ critical point that pass arbitrarily close to the $SO(2)_D\times SO(3)_{\textrm{diag}}$ $N=4$ critical point.         
\end{abstract}
\newpage
\section{Introduction}
The study of holographic RG flows has attracted much attention since the proposal of the AdS/CFT correspondence in \cite{maldacena}. These solutions take the form of domain walls interpolating between $AdS$ vacua or between an $AdS$ vacuum and a singular geometry and holographically descibe RG flows between conformal fixed points or from a conformal fixed point to a non-conformal phase in the dual field theories. In particular, holographic RG flows in AdS$_5$/CFT$_4$ correspondence provide a very useful tool to understand strongly coupled dynamics of four-dimensional gauge theories. A number of solutions describing various deformations of the $N=4$ Super Yang-Mills (SYM) theory, dual to type IIB theory on $AdS_5\times S^5$, has been constructed both in type IIB supergravity and in the efffective $N=8$ five-dimensional gauged supergravity, see for example \cite{GPPZ,FGPP,KW_5Dflow,N2_IIB_flow,AdS_RG_flow,N1_SYM_fixed_point}. The consistent truncation of type IIB supergravity on $S^5$ constructed recently in \cite{Henning_IIB_truncation} allows solutions in the $N=8$ gauged supergravity to be uplifted to ten dimensions.
\\
\indent On the other hand, similar solutions in gauged supergravities with $N<8$ supersymmetry are less known than those of the maximal case. In particular, a number of RG flow solutions in half-maximal $N=4$ gauged supergravity have appeared only very recently in \cite{5D_N4_flow_Davide} and \cite{5D_N4_flow}, see also \cite{Davide_5D_DW} for an eariler construction and \cite{5D_N2_DW1} for domain wall solutions in $N=2$ gauged supergravity. Furthermore, a number of RG flow solutions has been found in $N=4$ gauged supergravity coupled to three vector multiplets obtained from a consistent truncation of the maximal $SO(5)$ gauged supergravity in seven dimensions \cite{Calos_5D_N4_flow}. In this case, there is only one supersymmetric $AdS_5$ vacuum, and the solutions describe RG flows from a conformal fixed point to non-conformal phases dual to singular geometries in the IR. In this work, we hope to fill this gap by adding a new family of holographic RG flow solutions between conformal fixed points to these results. We will work in the half-maximal gauged supergravity with a new gauge group that leads to more interesting $AdS_5$ vacua and holographic RG flows, see \cite{7Dflow,7D_flow_Tomasiello,6Dflow,6Dflow1,tri-sasakian-flow,orbifold_flow,4Dflow,N4_Janus,N4_omega_flow,3Dflow} for similar studies in half-maximal gauged supergravity in other dimensions.
\\
 \indent $N=4$ gauged supergravity coupled to an arbitrary number $n$ of vector multiplets has $SO(1,1)\times SO(5,n)$ global symmetry. Gaugings of a subgroup $G_0\subset SO(1,1)\times SO(5,n)$ can be implemented by using the embedding tensor formalism \cite{N4_gauged_SUGRA}, see also \cite{5D_N4_Dallagata}. We will mainly consider the case of $n=5$ vector multiplets and $SO(2)_D\times SO(3)\times SO(3)\subset SO(5,5)$ gauge group. Unlike the previously studied $SO(2)\times SO(3)\times SO(3)$ gauge group in \cite{5D_N4_flow_Davide} and \cite{5D_N4_flow} in which the $SO(2)$ factor is a subgroup of the $SO(5)_R\sim USp(4)_R$ R-symmetry, the $SO(2)_D$ considered here is a diagonal subgroup of $SO(2)\subset SO(5)_R$ and $SO(2)\subset SO(5)$ symmetry of the vector multiplets. We will look for supersymmetric $AdS_5$ vacua dual to four-dimensional SCFTs and possible holographic RG flows between these SCFTs. 
\\ 
\indent Since the ungauged $N=4$ supergravity with $n=5$ vector multiplets can be obtained from a $T^5$ reduction of $N=1$ supergravity in ten dimensions, the $N=4$ gauged supergravity studied here might be possibly embedded in string/M-theory. However, it should be pointed out that this $N=4$ gauged supergravity has currently no known higher dimensional origin. Therefore, a complete holographic description is not available at this stage. However, even without higher-dimensional embedding, the results from the effective $N=4$ gauged supergravity in five dimensions could still be useful in a holographic study of strongly coupled four-dimensional SCFTs.
\\
\indent Similar studies in the cases of $SO(2)\times SO(3)\times SO(3)$ and $SO(2)_D\times SO(3)$ gauge groups have appeared recently in \cite{5D_N4_flow_Davide} and \cite{5D_N4_flow}. In the first case, there are two $N=4$ supersymmetric $AdS_5$ vacua while the second gauge group admits $N=4$ and $N=2$ $AdS_5$ vacua. The holographic RG flows in both cases have also been given in \cite{5D_N4_flow_Davide} and \cite{5D_N4_flow}. In the $SO(2)_D\times SO(3)\times SO(3)$ gauge group considered in this paper, we find a number of more interesting results. First of all, we discover four supersymmetric $AdS_5$ vacua within a truncation to $SO(2)_{\textrm{diag}}$ invariant sector of the $SO(1,1)\times SO(5,5)/SO(5)\times SO(5)$ scalar manifold. Two of the critical points preserve the maximal $N=4$ supersymmetry with $SO(2)_D\times SO(3)\times SO(3)$ and $SO(2)_D\times SO(3)_{\textrm{diag}}$ symmetries while the remaining two are only $N=2$ supersymmetric with $SO(2)_{\textrm{diag}}\times SO(3)$ and $SO(2)_{\textrm{diag}}$ symmetries. The first three critical points have an analogue in the results of \cite{5D_N4_flow_Davide} and \cite{5D_N4_flow} although with some differences in scalar mass spectra. The $N=2$ critical point with $SO(2)_{\textrm{diag}}$ symmetry is however entirely new.  
\\
\indent Another interesting result of the present paper is that there exists a family of RG flows from the $SO(2)_D\times SO(3)\times SO(3)$ $N=4$ critical point to the $SO(2)_{\textrm{diag}}$ $N=2$ critical point that consists of a direct flow and flows that pass arbitrarily close to the $SO(2)_D\times SO(3)_{\textrm{diag}}$ $N=4$ critical point. This is similar to the families of RG flows found in the maximal gauged supergravity in four dimensions \cite{Warner_4D_flow,Guarino_BPS_DW,Elec_mag_flows}. To the best of the author's knowledge, the present result is the first example of families, or cones in the terminology of \cite{Warner_4D_flow}, of RG flows in the framework of five-dimensional gauged supergravity. 
\\
\indent The paper is organized as follows. In section \ref{N4_SUGRA},
we review five-dimensional $N=4$ gauged supergravity coupled to vector multiplets in the embedding tensor formalism. In section \ref{result}, we consider the case of five vector multiplets and describe the embedding of $SO(2)_D\times SO(3)\times SO(3)$ gauge group in the $SO(5,5)$ global symmetry. Supersymmetric $AdS_5$ vacua and RG flows interpolating between these vacua are also given. Finally, we end the paper by giving some conclusions and comments in section \ref{conclusion}. 

\section{Five dimensional $N=4$ gauged supergravity coupled to vector multiplets}\label{N4_SUGRA} 
In this section, we give a brief review of five-dimensional $N=4$ gauged supergravity coupled to an arbitrary number $n$ of vector multiplets. We mainly focus on the scalar potential and supersymmetry transformations of fermions which are relevant for finding supersymmetric $AdS_5$ vacua and RG flow solutions. More details and the complete construction of $N=4$ gauged supergravity can be found in \cite{N4_gauged_SUGRA} and \cite{5D_N4_Dallagata}. 
\\
\indent The $N=4$ supergravity multiplet consists of the graviton
$e^{\hat{\mu}}_\mu$, four gravitini $\psi_{\mu i}$, six vectors $(A_\mu^0,A_\mu^m)$, four spin-$\frac{1}{2}$ fields $\chi_i$ and one real
scalar $\Sigma$, the dilaton. Space-time and tangent space indices are denoted respectively by $\mu,\nu,\ldots =0,1,2,3,4$ and
$\hat{\mu},\hat{\nu},\ldots=0,1,2,3,4$. The fundamental representation of $SO(5)_R\sim USp(4)_R$
R-symmetry is described by $m,n=1,\ldots, 5$ for the
$SO(5)_R$ and $i,j=1,2,3,4$ for the $USp(4)_R$. The latter also corresponds to spinor representation of $SO(5)_R$. For the $n$ vector multiplets, each multiplet contains a vector field $A_\mu$, four gaugini $\lambda_i$ and five scalars $\phi^m$. These multiplets will be labeled by indices $a,b=1,\ldots, n$. The corresponding component fields are accordingly denoted by $(A^a_\mu,\lambda^{a}_i,\phi^{ma})$. The $5n$ scalar fields parametrize the $SO(5,n)/SO(5)\times SO(n)$ coset.
\\
\indent Combining the gravity and vector multiplets, we have $6+n$ vector fields denoted collectively by $A^{\mc{M}}_\mu=(A^0_\mu,A^m_\mu,A^a_\mu)$ and $5n+1$ scalars in the $\mathbb{R}^+\times SO(5,n)/SO(5)\times SO(n)$ coset. All fermionic fields are symplectic Majorana spinors subject to the condition
\begin{equation}
\xi_i=\Omega_{ij}C(\bar{\xi}^j)^T 
\end{equation}
with $C$ and $\Omega_{ij}$ being the charge conjugation matrix and $USp(4)$ symplectic matrix, respectively.
\\
\indent The $SO(5,n)/SO(5)\times SO(n)$ coset can be described by a coset representative $\mc{V}_M^{\phantom{M}A}$ transforming under the global $G=SO(5,n)$ and the local $H=SO(5)\times SO(n)$ by left and right multiplications, respectively. We use the global $SO(5,n)$ indices $M,N,\ldots=1,2,\ldots , 5+n$. The local $H$ indices $A,B,\ldots$ can be split into $A=(m,a)$. The coset representative can then be written as
\begin{equation}
\mc{V}_M^{\phantom{M}A}=(\mc{V}_M^{\phantom{M}m},\mc{V}_M^{\phantom{M}a}).
\end{equation}
In addition, the matrix $\mc{V}_M^{\phantom{M}A}$ satisfies the relation
\begin{equation}
\eta_{MN}={\mc{V}_M}^A{\mc{V}_N}^B\eta_{AB}=-\mc{V}_M^{\phantom{M}m}\mc{V}_N^{\phantom{M}m}+\mc{V}_M^{\phantom{M}a}\mc{V}_N^{\phantom{M}a}
\end{equation}
with $\eta_{MN}=\textrm{diag}(-1,-1,-1,-1,-1,1,\ldots,1)$ being the $SO(5,n)$ invariant tensor. Furthermore, the $SO(5,n)/SO(5)\times SO(n)$ coset can also be described in terms of a symmetric matrix
\begin{equation}
M_{MN}=\mc{V}_M^{\phantom{M}m}\mc{V}_N^{\phantom{M}m}+\mc{V}_M^{\phantom{M}a}\mc{V}_N^{\phantom{M}a}
\end{equation}
which is manifestly $SO(5)\times SO(n)$ invariant.
\\
\indent Gaugings of $N=4$ supergravity can be efficiently described by using the embedding tensor which determines the embedding of admissible gauge groups in the global symmetry $SO(1,1)\times SO(5,n)$. Supersymmetry allows for three components of the embedding tensor of the form $\xi^{M}$, $\xi^{MN}=\xi^{[MN]}$ and $f_{MNP}=f_{[MNP]}$. These components also need to satisfy a set of quadratic constraints. Furthermore, the existence of supersymmetric $AdS_5$ vacua requires $\xi^M=0$ \cite{AdS5_N4_Jan}. Therefore, we will consider only gaugings with $\xi^{M}=0$. In this case, the gauge group is entirely embedded in $SO(5,n)$, and the quadratic constraints reduce to
\begin{equation}
f_{R[MN}{f_{PQ]}}^R=0\qquad \textrm{and}\qquad {\xi_M}^Qf_{QNP}=0\, .\label{QC}
\end{equation}
\indent With $\xi^{M}=0$, the gauge generators in the fundamental representation of $SO(5,n)$ can be written as
\begin{equation}
{(X_M)_N}^P=-{f_M}^{QR}{(t_{QR})_N}^P={f_{MN}}^P\quad \textrm{and}\quad {(X_0)_N}^P=-\xi^{QR}{(t_{QR})_N}^P={\xi_N}^P
\end{equation}
with ${(t_{MN})_P}^Q=\delta^Q_{[M}\eta_{N]P}$ being $SO(5,n)$ generators. The gauge covariant derivative reads
\begin{equation}
D_\mu=\nabla_\mu+A_\mu^{M}X_M+A^0_\mu X_0
\end{equation}
where $\nabla_\mu$ is the usual space-time covariant derivative. We use the definition of $\xi^{MN}$ and $f_{MNP}$ that includes the gauge coupling constants. Note also that $SO(5,n)$ indices $M,N,\ldots$ are lowered and raised by $\eta_{MN}$ and its inverse $\eta^{MN}$. 
\\
\indent In this paper, we are mainly interested in $AdS_5$ vacua and holographic RG flows in the form of domain walls that involve only the metric and scalar fields. For $AdS_5$ vacua with all scalars constant, this is awalys a consistent truncation. However, for holographic RG flows given by domain wall solutions with non-constant scalars, some of the scalars can be charged under the gauge group and couple to the gauge fields via the following Yang-Mills equation, see \cite{N4_gauged_SUGRA} for more detail,
\begin{equation} 
D_\mu (\Sigma^2 M_{MN}\mc{H}^{N\mu\nu})=\frac{1}{8}{X_{MP}}^RM_{RQ}D^\nu M^{PQ}\label{YM_eq}
\end{equation}
with the covariant field strengths defined by
\begin{equation}
\mc{H}^M_{\mu\nu}=2\pd_{[\mu}A^M_{\nu]}+{X_{NP}}^MA^N_\mu A^P_\nu+\frac{1}{2}\xi^{MN}B_{N\mu\nu}\, .
\end{equation}
It should be noted that the two-form fields are introduced in the embedding tensor formalism for the requirement of gauge covariance for the gauge field strengths. However, these are auxiliary fields since they do not have kinetic terms. In all the truncations we will consider here, it turns out that the Yang-Mills currents given by the right hand side of \eqref{YM_eq} vanish. Therefore, it is consistent to set all the gauge fields to zero in the RG flow solutions. With all vector fields vanishing, it is also consistent to set all the two-form fields to zero as can be seen from the corresponding field equation given in \cite{N4_gauged_SUGRA}. Accordingly, for simplicity of various expressions, we will set all the fields but the metric and scalar fields to zero from now on. 
\\
\indent  The bosonic Lagrangian of a general gauged $N=4$ supergravity coupled to $n$ vector multiplets can be written as
\begin{eqnarray}
e^{-1}\mc{L}&=&\frac{1}{2}R-\frac{3}{2}\Sigma^{-2}\pd_\mu \Sigma \pd^\mu \Sigma +\frac{1}{16} \pd_\mu M_{MN}\pd^\mu
M^{MN}-V\label{Lar}
\end{eqnarray}
where $e$ is the vielbein determinant. The scalar potential reads
\begin{eqnarray}
V&=&-\frac{1}{4}\left[f_{MNP}f_{QRS}\Sigma^{-2}\left(\frac{1}{12}M^{MQ}M^{NR}M^{PS}-\frac{1}{4}M^{MQ}\eta^{NR}\eta^{PS}\right.\right.\nonumber \\
& &\left.+\frac{1}{6}\eta^{MQ}\eta^{NR}\eta^{PS}\right) +\frac{1}{4}\xi_{MN}\xi_{PQ}\Sigma^4(M^{MP}M^{NQ}-\eta^{MP}\eta^{NQ})\nonumber \\
& &\left.
+\frac{\sqrt{2}}{3}f_{MNP}\xi_{QR}\Sigma M^{MNPQR}\right]
\end{eqnarray}
where $M^{MN}$ is the inverse of $M_{MN}$, and $M^{MNPQR}$ is obtained from
\begin{equation}
M_{MNPQR}=\epsilon_{mnpqr}\mc{V}_{M}^{\phantom{M}m}\mc{V}_{N}^{\phantom{M}n}
\mc{V}_{P}^{\phantom{M}p}\mc{V}_{Q}^{\phantom{M}q}\mc{V}_{R}^{\phantom{M}r}
\end{equation}
by raising indices with $\eta^{MN}$. 
\\
\indent Fermionic supersymmetry transformations are given by
\begin{eqnarray}
\delta\psi_{\mu i} &=&D_\mu \epsilon_i+\frac{i}{\sqrt{6}}\Omega_{ij}A^{jk}_1\gamma_\mu\epsilon_k,\\
\delta \chi_i &=&-\frac{\sqrt{3}}{2}i\Sigma^{-1} \pd_\mu
\Sigma\gamma^\mu \epsilon_i+\sqrt{2}\Omega_{ij}A_2^{kj}\epsilon_k,\\
\delta \lambda^a_i&=&i\Omega^{jk}({\mc{V}_M}^a\pd_\mu
{\mc{V}_{ij}}^M)\gamma^\mu\epsilon_k+\sqrt{2}\Omega_{ij}A_{2}^{akj}\epsilon_k\,
.
\end{eqnarray}
The fermion shift matrices are in turn defined by
\begin{eqnarray}
A_1^{ij}&=&-\frac{1}{\sqrt{6}}\left(\sqrt{2}\Sigma^2\Omega_{kl}{\mc{V}_M}^{ik}{\mc{V}_N}^{jl}\xi^{MN}+\frac{4}{3}\Sigma^{-1}{\mc{V}^{ik}}_M{\mc{V}^{jl}}_N{\mc{V}^P}_{kl}{f^{MN}}_P\right),\nonumber
\\
A_2^{ij}&=&\frac{1}{\sqrt{6}}\left(\sqrt{2}\Sigma^2\Omega_{kl}{\mc{V}_M}^{ik}{\mc{V}_N}^{jl}\xi^{MN}-\frac{2}{3}\Sigma^{-1}{\mc{V}^{ik}}_M{\mc{V}^{jl}}_N{\mc{V}^P}_{kl}{f^{MN}}_P\right),\nonumber
\\
A_2^{aij}&=&-\frac{1}{2}\left(\Sigma^2{{\mc{V}_M}^{ij}\mc{V}_N}^a\xi^{MN}-\sqrt{2}\Sigma^{-1}\Omega_{kl}{\mc{V}_M}^a{\mc{V}_N}^{ik}{\mc{V}_P}^{jl}f^{MNP}\right).
\end{eqnarray}
\indent $\mc{V}_M^{\phantom{M}ij}$ is defined in terms of ${\mc{V}_M}^m$ and $SO(5)$ gamma matrices ${\Gamma_{mi}}^j$ as
\begin{equation}
{\mc{V}_M}^{ij}=\frac{1}{2}{\mc{V}_M}^{m}\Gamma^{ij}_m
\end{equation}
with $\Gamma^{ij}_m=\Omega^{ik}{\Gamma_{mk}}^j$. Similarly, the inverse ${\mc{V}_{ij}}^M$ can be written as
\begin{equation}
{\mc{V}_{ij}}^M=\frac{1}{2}{\mc{V}_m}^M(\Gamma^{ij}_m)^*=\frac{1}{2}{\mc{V}_m}^M\Gamma_{m}^{kl}\Omega_{ki}\Omega_{lj}\,
.
\end{equation}
In this paper, we will use the following representation of $SO(5)$ gamma matrices
\begin{eqnarray}
\Gamma_1&=&-\sigma_2\otimes \sigma_2,\qquad \Gamma_2=i\mathbb{I}_2\otimes \sigma_1,\qquad \Gamma_3=\mathbb{I}_2\otimes \sigma_3,\nonumber\\
\Gamma_4&=&\sigma_1\otimes \sigma_2,\qquad \Gamma_5=\sigma_3\otimes \sigma_2
\end{eqnarray}
with $\sigma_i$, $i=1,2,3$, being the Pauli matrices.
\\
\indent The covariant derivative on $\epsilon_i$ is given by
\begin{equation}
D_\mu \epsilon_i=\pd_\mu \epsilon_i+\frac{1}{4}\omega_\mu^{ab}\gamma_{ab}\epsilon_i+{Q_{\mu i}}^j\epsilon_j
\end{equation}
with the composite connection defined by
\begin{equation}
{Q_{\mu i}}^j={\mc{V}_{ik}}^M\pd_\mu {\mc{V}_M}^{kj}\, .
\end{equation}
Finally, we note the relation between the scalar potential and the fermion shift matrices $A_1$ and $A_2$
\begin{equation}
V=-A^{ij}_1A_{1ij}+A^{ij}_2A_{2ij}+A_2^{aij}
A^a_{2ij}
\end{equation}
which is a consequence of the quadratic constraints \eqref{QC}. Note also that raising and lowering of $i,j,\ldots$ indices by $\Omega^{ij}$ and $\Omega_{ij}$ are related to complex conjugation.

\section{$N=4$ gauged supergravity with $SO(2)_D\times SO(3)\times SO(3)$ gauge group}\label{result}
We now consider $N=4$ gauged supergravity coupled to $n=5$ vector multiplets with the global symmetry group $SO(1,1)\times SO(5,5)$. We are interested in a compact gauge group of the form $SO(2)_D\times SO(3)\times SO(3)$ with the embedding tensor given by
\begin{eqnarray}
\xi^{MN}&=&g_1(\delta^M_1\delta^N_2-\delta^M_2\delta^N_1)-g_2(\delta^M_{10}\delta^N_9-\delta^M_9\delta^N_{10}),\\ 
f_{\tilde{m}+2,\tilde{n}+2,\tilde{p}+2}&=&h_1\epsilon_{\tilde{m}\tilde{n}\tilde{p}},\qquad \tilde{m},\tilde{n},\tilde{p}=1,2,3,\\
f_{\tilde{a}\tilde{b}\tilde{c}}&=&h_2\epsilon_{\tilde{a}\tilde{b}\tilde{c}},\qquad \tilde{a},\tilde{b},\tilde{c}=1,2,3
\end{eqnarray} 
where $g_1$, $g_2$, $h_1$ and $h_2$ are the corresponding coupling constants. 
\\
\indent The form of $\xi^{MN}$ implies that the $SO(2)_D$ is a diagonal subgroup of $SO(2)_R\subset SO(5)_R$ and $SO(2)\subset SO(5)$ generated by the $SO(5,5)$ generators $t_{12}$ and $t_{9,10}$. This factor is gauged by the vector field $A^0_\mu$ in the supergravity multiplet, see \cite{AdS5_N4_Jan} for more detail. Since $\xi_{MN}$ and $f_{MNP}$ have no indices in common, the second quadratic condition in \eqref{QC} is identically satisfied while the first condition holds by virtue of the Jacobi's identity for the two $SO(3)$ factors. Therefore, this is an admissible gauge group. Similar gauge groups with $g_2=0$ or $h_2=0$ have already been considered in \cite{5D_N4_flow_Davide} and \cite{5D_N4_flow}. 
\\
\indent To parametrize the coset representative for $SO(5,5)/SO(5)\times SO(5)$, we first identify the $SO(5,5)$ non-compact generators
\begin{equation}
Y_{ma}=t_{m,a+5},\qquad m=1,2,\ldots, 5,\qquad a=1,2,\ldots, 5\, .
\end{equation}
Dealing with all $25$ scalars in this coset is not practically possible, we will truncate to a smaller submanifold of $SO(5,5)/SO(5)\times SO(5)$ invariant under a certain subgroup of $SO(2)_D\times SO(3)\times SO(3)$. For later convenience, we will denote the gauge group by $SO(2)_D\times SO(3)_R\times SO(3)$ corresponding to the components $(\xi^{12}$, $\xi^{9,10})$, $f_{\tilde{m}\tilde{n}\tilde{p}}$ and $f_{\tilde{a}\tilde{b}\tilde{c}}$ of the embedding tensor, respectively. $SO(3)_R$ and $SO(3)$ are subgroups of the $SO(5)_R$ R-symmetry and the $SO(5)$ symmetry of the five vector multiplets. A simple truncation that is more manageable and still gives interesting results is given by scalars that are singlet under $SO(2)_{\textrm{diag}}\subset SO(2)_D\times SO(3)_R\times SO(3)$. In this truncation, $SO(2)_{\textrm{diag}}$, generated by a linear combination of $t_{12}$, $t_{45}$ and $t_{9,10}$, is the diagonal subgroup of $SO(2)_D\times SO(2)_R\times SO(2)$ with $SO(2)_R\times SO(2)\subset SO(3)_R\times SO(3)$. 
\\
\indent There are nine singlet scalars under $SO(2)_{\textrm{diag}}$ corresponding to the non-compact generators
\begin{eqnarray}
& &\hat{Y}_1=Y_{31},\qquad \hat{Y}_2=Y_{42}+Y_{53},\qquad \hat{Y}_3=Y_{44}+Y_{55},\nonumber \\
& &\hat{Y}_4=Y_{43}-Y_{52},\qquad \hat{Y}_5=Y_{45}-Y_{54},\qquad \hat{Y_6}=Y_{12}+Y_{23},\nonumber \\
& &\hat{Y}_7=Y_{13}-Y_{22},\qquad \hat{Y}_8=Y_{14}+Y_{25},\qquad \hat{Y}_9=Y_{15}-Y_{24}\, . 
\end{eqnarray}
The coset representative can then be written as
\begin{equation}
\mc{V}=e^{\phi_1\hat{Y}_1}e^{\phi_2\hat{Y}_2}e^{\phi_3\hat{Y}_3}e^{\phi_4\hat{Y}_4}e^{\phi_5\hat{Y}_5}
e^{\phi_6\hat{Y}_6}e^{\phi_7\hat{Y}_7}e^{\phi_8\hat{Y}_8}e^{\phi_9\hat{Y}_9}\, .\label{coset_rep}
\end{equation}
It turns out that the resulting scalar potential and fermion shift matrices computed from this coset representative are still highly complicated. However, it can be straightforwardly verified that setting $\phi_i=0$, for $i=4,\ldots, 9$, is a consistent truncation. In this case, the analysis simplifies considerably. In particular, the $A^{ij}_1$ tensor is diagonal in this subtruncation. Furthermore, non-vanishing $\phi_i$, $i=4,\ldots, 9$, do not give rise to any new $AdS_5$ vacua other than those given below. We then make a subtruncation by setting $\phi_i=0$, $i=4,\ldots, 9$, for simplicity.
\\
\indent Using \eqref{coset_rep} with $\phi_4=\ldots =\phi_9=0$, we find the scalar potential
\begin{eqnarray}
V&=&\frac{1}{8}g_2^2\sinh^22\phi_3\Sigma^4+\frac{1}{8}\Sigma^{-2}\cosh^2\phi_3\left[2h_1h_2\cosh^2\phi_3\sinh2\phi_1\sinh^22\phi_2 \right.\nonumber \\
& &+h_1^2\cosh^2\phi_2\left\{[\cosh2\phi_1(\cosh2\phi_2-3)-4]+2\cosh2\phi_1\cosh^2\phi_2\cosh2\phi_3\right\}\nonumber \\
& &\left.+h_2^2\left\{\sinh^2\phi_2[\cosh2\phi_1(3+\cosh2\phi_2)-4]+2\cosh2\phi_1\cosh2\phi_3\sinh^4\phi_2 \right\}\right]\nonumber \\
& &+\sqrt{2}g_1\cosh^2\phi_3\left(h_1\cosh\phi_1\cosh^2\phi_2+h_2\sinh\phi_1\sinh^2\phi_2\right)\Sigma\, .\label{poten}
\end{eqnarray}
The $A_1^{ij}$ tensor is given by
\begin{equation}
A^{ij}_1=\textrm{diag}(\alpha_1,\alpha_2,\alpha_1,\alpha_2)
\end{equation}
with
\begin{eqnarray}
\alpha_{1,2}&=&\pm \frac{1}{\sqrt{6}}\Sigma^{-1}\cosh^2\phi_3\left(h_1\cosh\phi_1\cosh^2\phi_2+h_2\sinh\phi_1\sinh^2\phi_2\right)\nonumber \\
& &+\frac{1}{4\sqrt{3}}(g_2\mp 2g_1-g_2\cosh2\phi_3)\Sigma^2\, .
\end{eqnarray}
Either of $\alpha_1$ or $\alpha_2$ leads to the superpotential in terms of which the scalar potential can be written as
\begin{equation}
V=\frac{3}{2}\Sigma^{2}\left(\frac{\pd W}{\pd \Sigma}\right)^2+\frac{9}{2}\left(\frac{\pd W}{\pd \phi_1}\right)^2+\frac{9}{4}\textrm{sech}^{2}\phi_3\left(\frac{\pd W}{\pd \phi_2}\right)^2+\frac{9}{4}\left(\frac{\pd W}{\pd \phi_3}\right)^2-6W^2
\end{equation}
for $W=W_1=\sqrt{\frac{2}{3}}|\alpha_{1}|$ or $W=W_2=\sqrt{\frac{2}{3}}|\alpha_{2}|$.
\\
\indent In this paper, we are only intested in supersymmetric $AdS_5$ vacua which are critical points of both the scalar potential and the superpotential with negative cosmological constants. Before giving these critical points, we first note that in general, the two eigenvalues of $A^{ij}_1$ are not equal but related by
\begin{equation}
\alpha_1+\alpha_2=-\frac{g_2}{\sqrt{3}}\Sigma^2\sinh^2\phi_3\, .
\end{equation}
We see that either $W_1$ or $W_2$ corresponds to $N=2$ supersymmetry but for $g_2=0$ or $\phi_3=0$, we find, with $\alpha_1=-\alpha_2$, $W_1=W_2$ leading to $N=4$ supersymmetry as studied in \cite{5D_N4_flow_Davide} and \cite{5D_N4_flow}. In particular, for $g_2\neq 0$, non-vanishing $\phi_3$ breaks supersymmetry from $N=4$ to $N=2$. 

\subsection{Supersymmetric $AdS_5$ vacua}
From the scalar potential \eqref{poten}, there are four supersymmetric $AdS_5$ critical points:
\begin{itemize}
\item I. The first critical point is given by
\begin{equation}
\phi_1=\phi_2=\phi_3=0,\qquad \Sigma=-\left(\frac{h_1}{\sqrt{2}g_1}\right)^{\frac{1}{3}},\qquad V_0=-3\left(\frac{g_1h_1^2}{2}\right)^{\frac{2}{3}}
\end{equation}
with $V_0$ being the value of the scalar potential at the critical point or the cosmological constant. This critical point preserves the full $SO(2)_D\times SO(3)\times SO(3)$ gauge symmetry and $N=4$ supersymmetry due to the vanishing $\phi_3$. We can rescale the dilaton $\Sigma$ or equivalently set $g_1=-\frac{h_1}{\sqrt{2}}$ to bring the critical point to $\Sigma=1$. Accordingly, this critical point is located at the origin of the $SO(1,1)\times SO(5,5)/SO(5)\times SO(5)$ scalar manifold and usually referred to as the trivial $AdS_5$ critical point.

\item II. There is another $N=4$ $AdS_5$ critical point given by
\begin{eqnarray}
& &\phi_1=\pm\phi_2=\frac{1}{2}\ln\left[\frac{h_2-h_1}{h_2+h_1}\right],\qquad \Sigma=-\left(\frac{h_1h_2}{\sqrt{2}g_1\sqrt{h_2^2-h_1^2}}\right)^{\frac{1}{3}},\nonumber \\
& &\phi_3=0,\qquad V_0=-\frac{3}{2}\left(\frac{\sqrt{2}g_1h_1^2h_2^2}{h_2^2-h_1^2}\right)^{\frac{2}{3}}\, .
\end{eqnarray}
This critical point preserves $SO(2)_D\times SO(3)_{\textrm{diag}}$ symmetry. Both signs of $\phi_2$ lead to equivalent critical points with the same cosmological constant and scalar masses.

\item III. The next critical point is given by
\begin{eqnarray}
& &\phi_1=\phi_2=0,\qquad \phi_3=\frac{1}{2}\ln\left[\frac{g_2-4g_1\pm 2\sqrt{4g_1^2-2g_1g_2-2g_2^2}}{3g_2}\right],\nonumber \\
& & \Sigma=-\left(\frac{\sqrt{2}h_1}{g_2}\right)^{\frac{1}{3}},\qquad V_0=-\frac{1}{3}(g_1-g_2)^2\left(\frac{\sqrt{2}h_1}{g_2}\right)^{\frac{4}{3}}\, .\label{PointIII}
\end{eqnarray}
Since $\phi_3\neq0$, this critical point is $N=2$ supersymmetric with $SO(2)_{\textrm{diag}}\times SO(3)$ symmetry. $SO(2)_{\textrm{diag}}$ is the diagonal subgroup of $SO(2)_R\times SO(2)_D$ with $SO(2)_R\subset SO(3)_R$. The two sign choices in \eqref{PointIII} give equivalent critical points.

\item IV. The last critical point is also $N=2$ supersymmetric given by
\begin{eqnarray}
& &\phi_1=\pm\phi_2=\frac{1}{2}\ln\left[\frac{h_2-h_1}{h_2+h_1}\right],\qquad \Sigma=\left(\frac{\sqrt{2}h_1h_2}{g_2\sqrt{h_2^2-h_1^2}}\right)^{\frac{1}{3}},\nonumber \\
& &\phi_3=\frac{1}{2}\ln\left[\frac{g_2-4g_1\pm 2\sqrt{4g_1^2-2g_1g_2-2g_2^2}}{3g_2}\right],\nonumber \\
& & V_0=-\frac{2}{3}(g_1-g_2)^2\left(\frac{h_1^2h_2^2}{\sqrt{2}g_2^2(h_2^2-h_1^2)}\right)^{\frac{2}{3}}\, .
\end{eqnarray}
At this critical point, the gauge group is broken down to just $SO(2)_{\textrm{diag}}$ with $SO(2)_{\textrm{diag}}$ being the diagonal subgroup of $SO(2)_D\times SO(2)_R\times SO(2)$. As in critical point III, the four sign choices lead to equivalent critical points.
 \end{itemize} 
We note that there are various possible values of the coupling constants $h_1$ and $h_2$ in order for critical points II and IV to exist. For definiteness, we will choose both $h_1$ and $h_2$ to be positive and take $h_2>h_1$. We will also choose the upper sign choice for critical points II, III and IV.
\\  
\indent Scalar masses at these critical points are given in table \ref{table1} to \ref{table4}. To simplify the expressions, we have followed the notation of \cite{5D_N4_flow_Davide} by redefining the coupling constants as
\begin{equation} 
g_1=-\frac{g}{\sqrt{2}},\qquad h_1=g,\qquad g_2=\frac{\sqrt{2}g}{\rho}\, .\label{g_redef}
\end{equation}
We also note that the existence of the $N=2$ $AdS_5$ critical point III requires $\rho>1$ as pointed out in \cite{5D_N4_flow_Davide}. In the tables, we have given conformal dimensions of the operators dual to the scalar fields of $N=4$ gauged supergravity by using the relation $m^2L^2=\Delta(\Delta-4)$ with the $AdS_5$ radius given by $L^2=-\frac{6}{V_0}$. For some values of $m^2L^2$, we have given only one root of $\Delta$ since the other root violates the unitarity bound $\Delta>1$ for all values of $\rho>1$. 
\\
\indent For $g_2=0$ or $\rho\rightarrow \infty$, scalar masses for critical points I and II reduce exactly to those given in \cite{5D_N4_flow} for $SO(2)\times SO(3)\times SO(3)$ gauge group. Note that all values of $m^2L^2$ and $\Delta$ do not depend on the $SO(3)$ coupling constant $h_2$. In the dual SCFTs, this $SO(3)\subset SO(5)$, symmetry of the vector multiplets, corresponds to a flavor symmetry. We also point out that the conformal dimensions $\Delta=-4,2+\frac{2}{\rho}$ and $\Delta= 3+\sqrt{25-\frac{72}{2+\rho}}, 1+\sqrt{25-\frac{72}{2+\rho}}$ at critical points I and III are in agreement with those of the two scalar modes considered in \cite{5D_N4_flow_Davide}. However, the full scalar masses have not been given in \cite{5D_N4_flow_Davide}. We hope the present results fill this gap and could be useful in other holographic study. 
\\
\indent Three of the eight massless scalars at critical point II are Goldstone bosons for the symmetry breaking $SO(2)_D\times SO(3)\times SO(3)$ to $SO(2)_D\times SO(3)_{\textrm{diag}}$. The remaining five massless scalars correspond to marginal deformations of the dual $N=2$ SCFT. Similarly, three and six massless scalars at critical points III and IV are all Goldstone bosons corresponding to the symmetry breaking $SO(2)_D\times SO(3)\times SO(3)\rightarrow SO(2)_{\textrm{diag}}\times SO(3)$ and $SO(2)_D\times SO(3)\times SO(3)\rightarrow SO(2)_{\textrm{diag}}$, respectively. However, in these cases, there are no marginal deformations in the dual $N=1$ SCFTs. 

\begin{table}[h]
\centering
\begin{tabular}{|c|c|}
  \hline
 $m^2L^2\phantom{\frac{1}{2}}$ & $\Delta$  \\ \hline
    $-4_{\times 10}$ &  $2$  \\
   $-3_{\times 6}$ &  $3$  \\
    $-4\left(1-\frac{1}{\rho^2}\right)_{\times 6}$ &  $2+\frac{2}{\rho}$  \\
 $\left( -3+\frac{4}{\rho}+\frac{4}{\rho^2}\right)_{\times 2}$ &  $3+\frac{2}{\rho}$  \\
 $\left( -3-\frac{4}{\rho}+\frac{4}{\rho^2}\right)_{\times 2}$ &  $3-\frac{2}{\rho}$, $1+\frac{2}{\rho}$  \\
  \hline
\end{tabular}
\caption{Scalar masses at the $N=4$ supersymmetric $AdS_5$ critical
point I with $SO(2)_D\times SO(3)\times SO(3)$ symmetry and the
corresponding dimensions of the dual operators.}\label{table1}
\end{table}

\begin{table}[h]
\centering
\begin{tabular}{|c|c|c|}
  \hline
   $m^2L^2\phantom{\frac{1}{2}}$ & $\Delta$  \\ \hline
   $0_{\times 8}$ &  $4$  \\
   $-4$ &  $2$  \\
    $5_{\times 6}$ &  $5$  \\
 $12$ &  $6$  \\
 $-4\left(1-\frac{1}{\rho^2}\right)_{\times 6}$ &  $2+\frac{2}{\rho}$  \\
 $\left( -3+\frac{4}{\rho}+\frac{4}{\rho^2}\right)_{\times 2}$ &  $3+\frac{2}{\rho}$  \\
 $\left( -3-\frac{4}{\rho}+\frac{4}{\rho^2}\right)_{\times 2}$ &  $3-\frac{2}{\rho}$, $1+\frac{2}{\rho}$  \\
  \hline
\end{tabular}
\caption{Scalar masses at the $N=4$ supersymmetric $AdS_5$ critical
point II with $SO(2)_D\times SO(3)_{\textrm{diag}}$ symmetry and the
corresponding dimensions of the dual operators.}\label{table2}
\end{table}

\begin{table}[h]
\centering
\begin{tabular}{|c|c|}
  \hline
   $m^2L^2\phantom{\frac{1}{2}}$ & $\Delta$  \\ \hline
    $0_{\times 3}$ &  $4$  \\
   $-4_{\times 3}$ &  $2$  \\
    $\left. -\frac{3\rho(8+\rho)}{(2+\rho)^2}\right|_{\times 6}$ &  $\frac{3\rho}{2+\rho}$, $\frac{8+\rho}{2+\rho}$  \\
 $\left.-\frac{12(1+2\rho)}{(2+\rho)^2}\right|_{\times 6}$ &  $\frac{6}{2+\rho}$, $\frac{2+4\rho}{2+\rho}$  \\
 $\left.-\frac{48(\rho-1)}{(2+\rho)^2}\right|_{\times 2}$ &  $\frac{12}{2+\rho}$, $\frac{4(\rho-1)}{2+\rho}$  \\
 $\left.\frac{3(7\rho-10)}{2+\rho}\right|_{\times 2}$ &  $2+\sqrt{\frac{25\rho-22}{2+\rho}}$ \\ 
 $\left.\frac{3(28-12\rho-\rho^2)}{(2+\rho)^2}\right|_{\times 2}$ &  $\frac{3(\rho-2)}{2+\rho}$, $\frac{14+\rho}{2+\rho}$  \\  
$22-\frac{72}{2+\rho}-2\sqrt{25-\frac{72}{2+\rho}}$ & $1+\sqrt{25-\frac{72}{2+\rho}}$ \\   
$22-\frac{72}{2+\rho}+2\sqrt{25-\frac{72}{2+\rho}}$ & $3+\sqrt{25-\frac{72}{2+\rho}}$ \\      
       \hline
\end{tabular}
\caption{Scalar masses at the $N=2$ supersymmetric $AdS_5$ critical
point III with $SO(2)_{\textrm{diag}}\times SO(3)$ symmetry and the
corresponding dimensions of the dual operators.}\label{table3}
\end{table}

\begin{table}[h]
\centering
\begin{tabular}{|c|c|c|}
  \hline
   $m^2L^2\phantom{\frac{1}{2}}$ & $\Delta$  \\ \hline
   $0_{\times 6}$ &  $4$  \\
   $\left. -\frac{3(\rho^2-16)}{(2+\rho)^2}\right|_{\times 2}$ &  $\frac{3(\rho+4)}{2+\rho}$  \\
    $\left. -\frac{4(\rho-1)(\rho+5)}{(2+\rho)^2}\right|_{\times 2}$ &  $\frac{2(5+\rho)}{2+\rho}$, $\frac{2(\rho-1)}{2+\rho}$  \\
 $\left.-\frac{48(\rho-1)}{(2+\rho)^2}\right|_{\times 4}$ &  $\frac{12}{2+\rho}$, $\frac{4(\rho-1)}{2+\rho}$  \\
 $\left.-\frac{3(\rho-6)}{2+\rho}\right|_{\times 2}$ &  $2+\sqrt{\frac{\rho+26}{2+\rho}}$  \\
 $\left.\frac{3(7\rho-10)}{2+\rho}\right|_{\times 2}$ &  $2+\sqrt{\frac{25\rho-22}{2+\rho}}$  \\
 $\left.-\frac{3(\rho^2+12\rho-28)}{(2+\rho)^2}\right|_{\times 4}$ &  $\frac{\rho+14}{2+\rho}$, $\frac{3(\rho-2)}{2+\rho}$  \\
 $-2+\frac{24}{2+\rho}+2\sqrt{\frac{26+\rho}{2+\rho}}$ & $3+\sqrt{\frac{26+\rho}{2+\rho}}$\\
 $-2+\frac{24}{2+\rho}-2\sqrt{\frac{26+\rho}{2+\rho}}$ & $1+\sqrt{\frac{26+\rho}{2+\rho}}$\\ 
 $22-\frac{72}{2+\rho}-2\sqrt{25-\frac{72}{2+\rho}}$ & $1+\sqrt{25-\frac{72}{2+\rho}}$ \\   
$22-\frac{72}{2+\rho}+2\sqrt{25-\frac{72}{2+\rho}}$ & $3+\sqrt{25-\frac{72}{2+\rho}}$ \\
  \hline
\end{tabular}
\caption{Scalar masses at the $N=2$ supersymmetric $AdS_5$ critical
point IV with $SO(2)_{\textrm{diag}}$ symmetry and the
corresponding dimensions of the dual operators.}\label{table4}
\end{table}

\subsection{Holographic RG flows}
\begin{figure}
         \centering
         \begin{subfigure}[b]{0.45\textwidth}
                 \includegraphics[width=\textwidth]{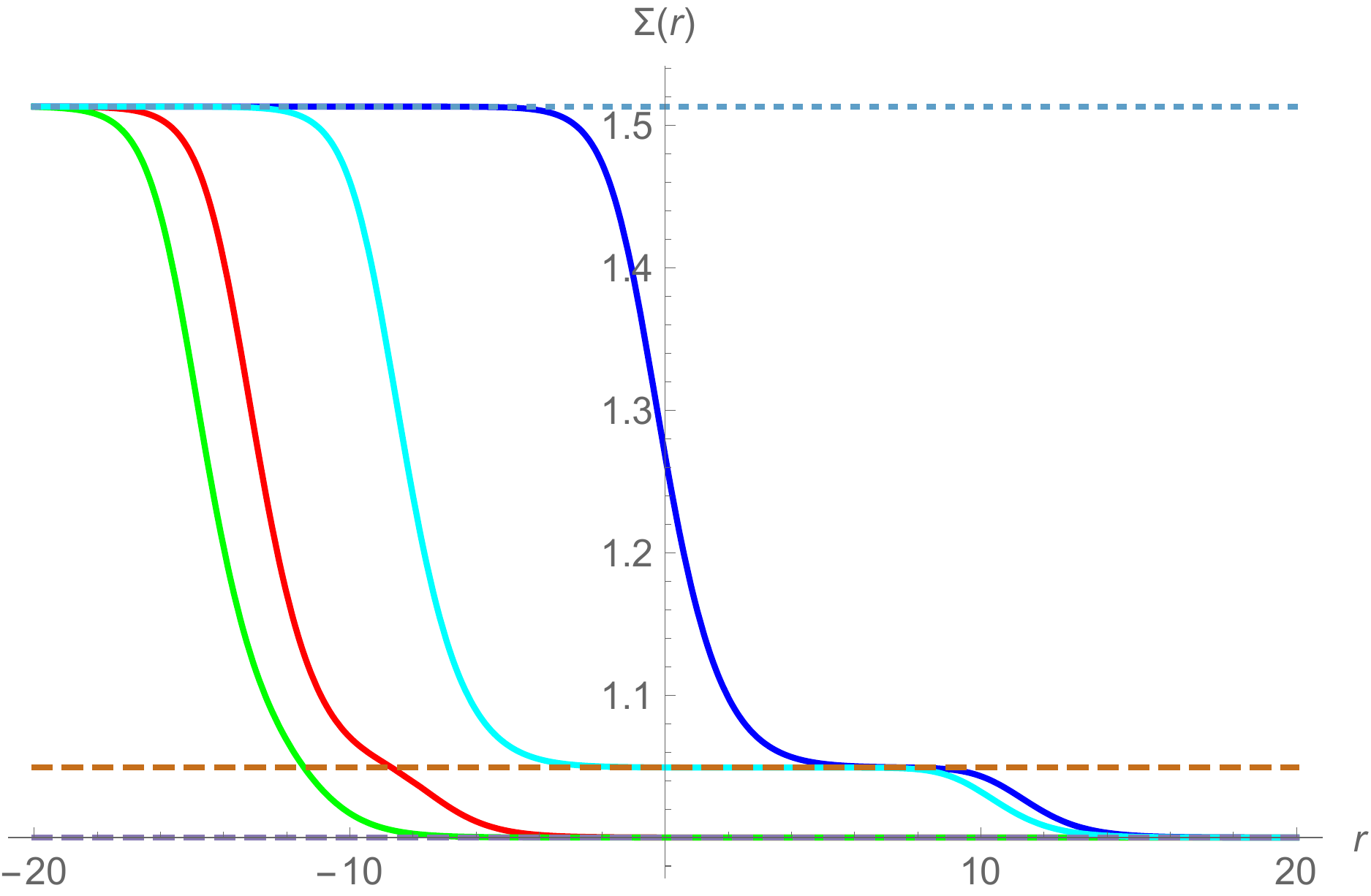}
                 \caption{Solution for $\Sigma(r)$}
         \end{subfigure} \qquad
\begin{subfigure}[b]{0.45\textwidth}
                 \includegraphics[width=\textwidth]{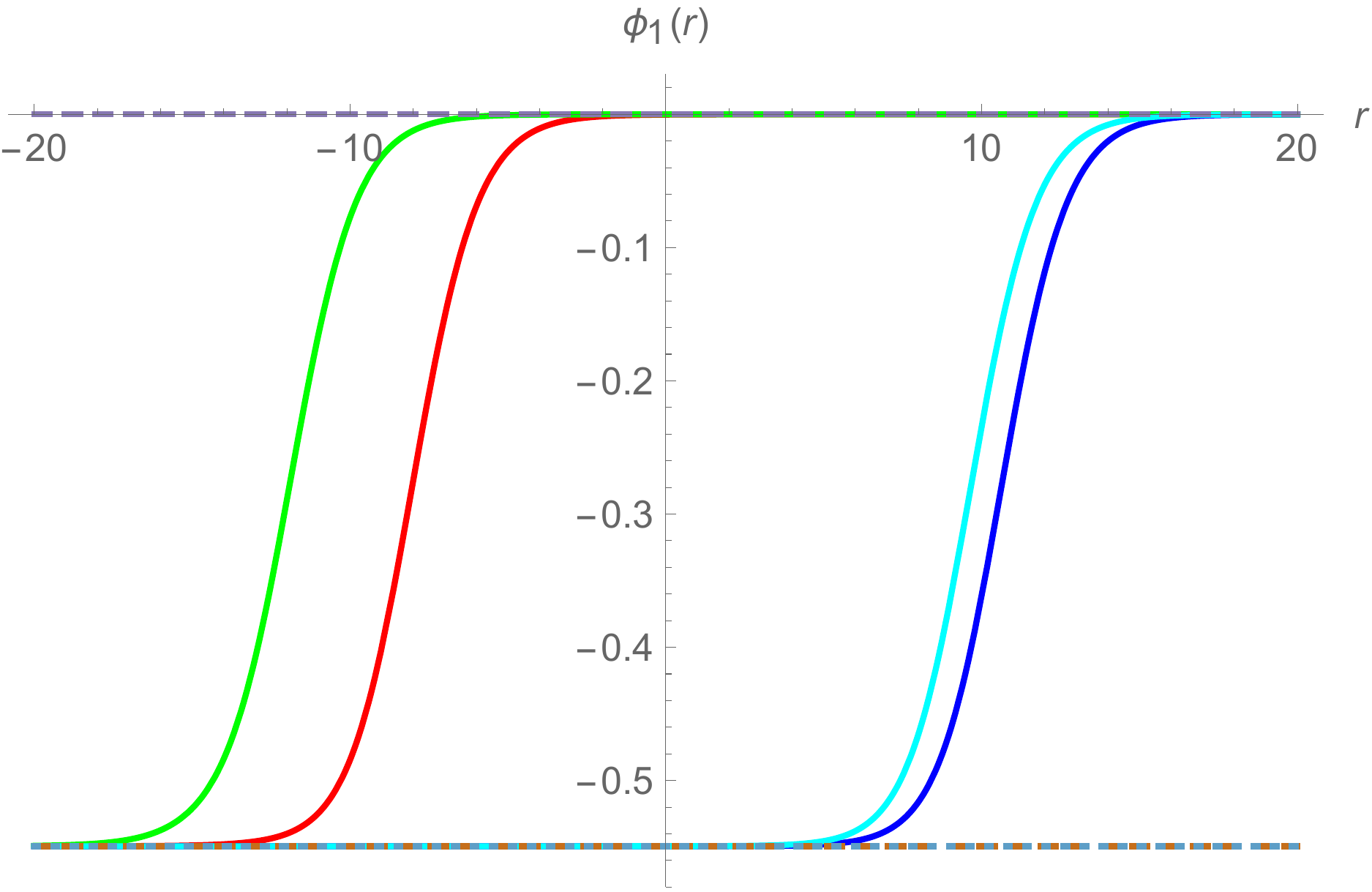}
                 \caption{Solution for $\phi_1(r)$}
         \end{subfigure}\\
         \begin{subfigure}[b]{0.45\textwidth}
                 \includegraphics[width=\textwidth]{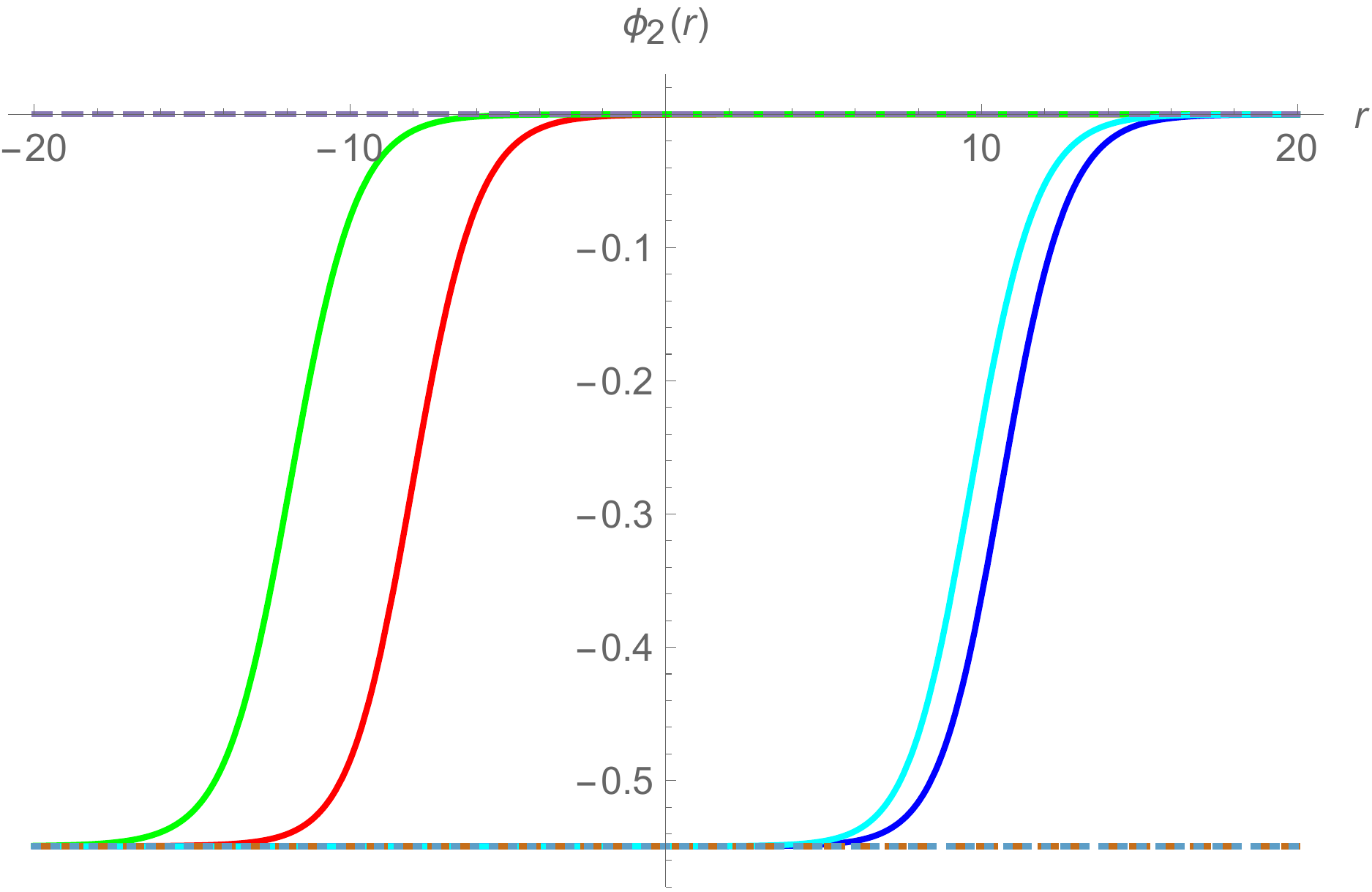}
                 \caption{Solution for $\phi_2(r)$}
         \end{subfigure}\qquad 
         \begin{subfigure}[b]{0.45\textwidth}
                 \includegraphics[width=\textwidth]{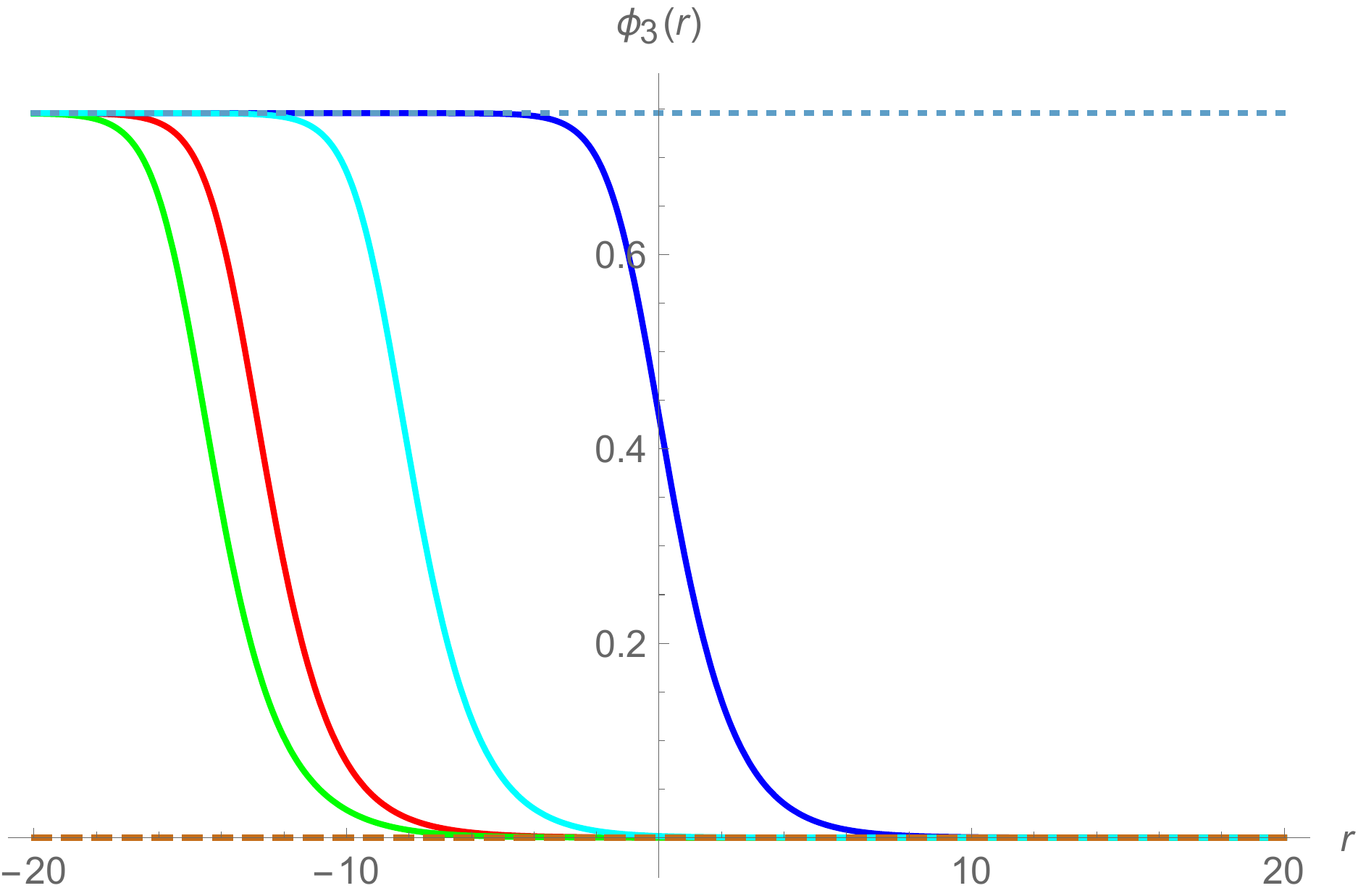}
                 \caption{Solution for $\phi_3(r)$}
         \end{subfigure}\\
          \begin{subfigure}[b]{0.45\textwidth}
                 \includegraphics[width=\textwidth]{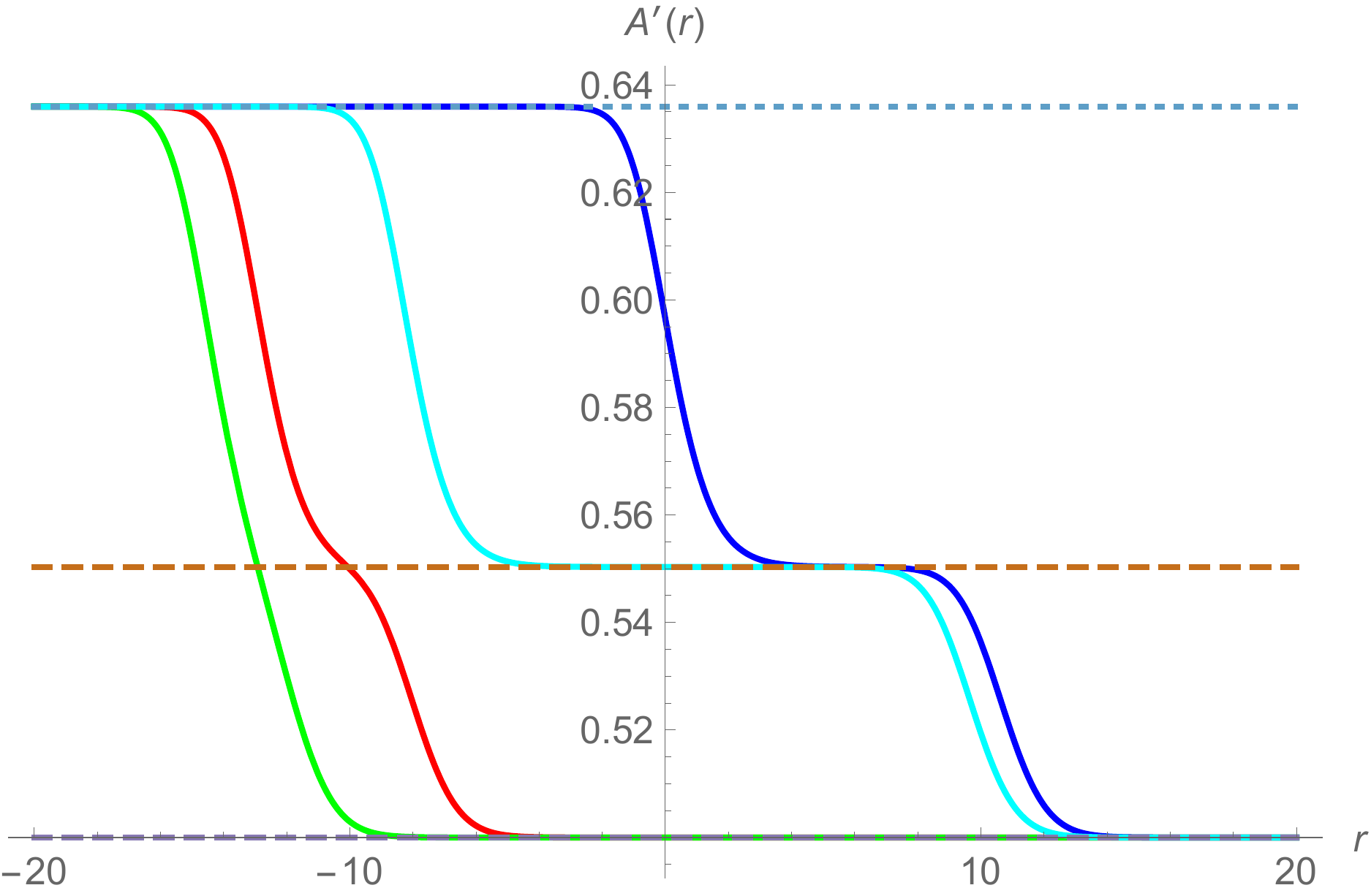}
                 \caption{Solution for $A'(r)$}
         \end{subfigure} 
         \caption{A family of RG flows from $N=4$ $AdS_5$ critical point with $SO(2)_D\times SO(3)\times SO(3)$ symmetry to $N=2$ $AdS_5$ fixed point with $SO(2)_{\textrm{diag}}$ symmetry in the IR with $g=1$, $\rho=3$ and $\zeta=\frac{1}{2}$.}\label{fig1}
 \end{figure}

We now look for holographic RG flow solutions interpolating between supersymmetric $AdS_5$ vacua identified in the previous section. These solutions take the form of domain walls in $N=4$ gauged supergravity described by the metric ansatz    
\begin{equation}
ds^2=e^{2A(r)}dx^2_{1,3}+dr^2\label{metric}
\end{equation} 
with $dx^2_{1,3}=\eta_{\alpha\beta}dx^\alpha dx^\beta$, $\alpha,\beta=0,1,2,3$, being the four-dimensional Minkowski metric. All the scalar fields and the Killing spinors $\epsilon_i$ are functions of only the radial coordinate $r$.
\\
\indent Before analysing the BPS equations, we first note the scalar kinetic terms
\begin{equation}
\mc{L}_{\textrm{kin}}=-\frac{3}{2}\Sigma^{-2}{\Sigma'}^2-\frac{1}{2}{\phi_1'}^2-\cosh^2\phi_3{\phi_2'}^2-{\phi_3'}^2
\end{equation}
in which we have denoted the $r$-derivatives by $'$. We now consider supersymmetry transformations of $\psi_{\mu i}$, $\chi_i$ and $\lambda^a_i$. By setting these to zero with the metric ansatz \eqref{metric}, we obtain the corresponding BPS equations for the RG flow solutions of interest. The analysis is essentially the same as in \cite{5D_N4_flow_Davide} and \cite{5D_N4_flow}, so we will mainly give the results with some detail omitted.
\\
\indent From the $\delta \psi_{\hat{\alpha} i}=0$ conditions, we find
\begin{equation}
A'\gamma_r\epsilon_i+i\sqrt{\frac{2}{3}}\Omega_{ij}A^{jk}_1\epsilon_k=0\, .\label{psi_eq1}
\end{equation}
Multiply by $A'\gamma_r$ and use \eqref{psi_eq1} again, we find
\begin{equation}
A'^2\epsilon_i+{M_i}^k{M_k}^j\epsilon_j=0\label{psi_eq2}
\end{equation}
for ${M_i}^j=\sqrt{\frac{2}{3}}\Omega_{ik}A_1^{kj}$. Non-vanishing solutions for $\epsilon_i$ implies that ${M_i}^k{M_k}^j\propto \delta^j_i$. We then write
\begin{equation}
{M_i}^k{M_k}^j=-|W|^2\delta^j_i
\end{equation}
where $W$ is the superpotential identified with the eigenvalues $\alpha_1$ or $\alpha_2$ of the $A^{ij}_1$ tensor. Recall that choosing one of these eigenvalues breaks half of the $N=4$ supersymmetry. By choosing $W=\sqrt{\frac{2}{3}}\alpha_1$, we find that the corresponding Killing spinors are given by $\epsilon_{1,3}$ while supersymmetry associated with $\epsilon_{2,4}$ is broken. We then set $\epsilon_2=\epsilon_4=0$ or equivalently impose the following projector
\begin{equation}
{\mathbb{P}_i}^j\epsilon_j=\epsilon_i\label{N2_Pro}
\end{equation}
for $\mathbb{P}=\textrm{diag}(1,0,1,0)$.
\\
\indent Using all these results in equations \eqref{psi_eq1} and \eqref{psi_eq2}, we find the flow equation for the metric function
\begin{equation}
A'=\pm |W|\, .\label{Ap_eq}
\end{equation}
together with the $\gamma^r$-projector
\begin{equation}
\gamma_r\epsilon_i=\mp i {I_i}^j\epsilon_j\label{gamma_r_projector}
\end{equation} 
with ${I_i}^j$ defined by
\begin{equation}
{I_i}^j=\sqrt{\frac{2}{3}}\frac{\Omega_{ik}A_1^{kj}}{|W|}\, .
\end{equation}
It should be noted that for $\phi_3=0$, we have $\alpha_1=-\alpha_2$ leading to $N=4$ supersymmetry. In this case, the projector \eqref{N2_Pro} is not needed.
\\
\indent As expected, the condition $\delta \psi_{\hat{r}i}=0$ gives the $r$-dependent Killing spinors of the form $\epsilon_i=e^{\frac{A}{2}}\epsilon_{0i}$ for constant spinors $\epsilon_{0i}$ satisfying \eqref{N2_Pro} and \eqref{gamma_r_projector}. Finally, using the projector \eqref{gamma_r_projector} in $\delta\chi_i=0$ and $\delta \lambda^a_i=0$ equations, we find the first-order flow equations for scalar fields. 
\\
\indent By the procedure described above, we obtain the following BPS equations
\begin{eqnarray}
\Sigma'&=&\frac{1}{3}\cosh^2\phi_3\left(h_1\cosh\phi_1\cosh^2\phi_2+h_2\sinh\phi_1\sinh^2\phi_2\right)\nonumber \\
& &+\frac{\sqrt{2}}{6}\Sigma^3(2g_1-g_2+g_2\cosh2\phi_3),\\
\phi_1'&=&-\Sigma^{-1}\cosh^2\phi_3\left(h_1\cosh^2\phi_2\sinh\phi_1+h_2\cosh\phi_1\sinh^2\phi_2\right),\\
\phi_2'&=&-\Sigma^{-1}\cosh\phi_2\sinh\phi_2(h_1\cosh\phi_1+h_2\sinh\phi_1),\\
\phi_3'&=&-\frac{1}{2}\Sigma^{-1}\sinh2\phi_3\left(h_1\cosh\phi_1\cosh^2\phi_2+h_2\sinh\phi_1\sinh^2\phi_2\right)\nonumber \\
& &+\frac{\sqrt{2}}{4}g_2\Sigma^2\sinh2\phi_3,\\
A'&=&\frac{1}{3}\Sigma^{-1}\cosh^2\phi_3\left(h_1\cosh\phi_1\cosh^2\phi_2+h_2\sinh\phi_1\sinh^2\phi_2\right)\nonumber \\
& &+\frac{\sqrt{2}}{12}\Sigma^2(g_2-2g_1-g_2\cosh2\phi_3)
\end{eqnarray}
which can be rewritten more compactly in terms of the superpotential as 
\begin{eqnarray}
& &\Sigma'=-\Sigma^2\frac{\pd W}{\pd \Sigma},\qquad \phi_1'=-3\frac{\pd W}{\pd \phi_1},\nonumber \\
 & &\phi_2'=-\frac{3}{2}\textrm{sech}^2\phi_3\frac{\pd W}{\pd \phi_2},\qquad \phi_3'=-\frac{3}{2}\frac{\pd W}{\pd \phi_3},\qquad A'=W\, .
  \end{eqnarray}
In deriving these equations we have chosen the upper sign choice in \eqref{Ap_eq} and \eqref{gamma_r_projector}. This also allows for identifying the UV and IR fixed points as the $AdS_5$ critical points at $r\rightarrow \infty$ and $r\rightarrow -\infty$, respectively. It can also be verified that the BPS equations are compatible with the second-order field equations obtained from the Lagrangian \eqref{Lar}. For $g_2=0$, the BPS equations reduce to those of $N=4$ RG flows studied in \cite{5D_N4_flow_Davide} and \cite{5D_N4_flow} while for $h_2=0$, we recover the BPS equations for $N=2$ RG flows in \cite{5D_N4_flow_Davide}. The former describes holographic RG flows between $N=4$ critical points I and II and can be obtained analytically. The latter corresponding to holographic RG flows from an $N=2$ SCFT in the UV to an $N=1$ SCFT in the IR have been obtained numerically. 
\\
\indent Although the $N=4$ $AdS_5$ vacua in the $SO(2)_D\times SO(3)\times SO(3)$ gauge group considered here have some of the scalar masses different from those in the $SO(2)\times SO(3)\times SO(3)$ gauge group studied in \cite{5D_N4_flow_Davide} and \cite{5D_N4_flow} as seen from table \ref{table1}, it turns out that setting $\phi_3=0$ and $\phi_1=\phi_2$ leads to the same BPS equations as those in \cite{5D_N4_flow_Davide} and \cite{5D_N4_flow}. Therefore, the flow solutions are the same and will not be repeated here. For other possible RG flows, we are not able to find any analytic solutions. Accordingly, we will rely on a numerical analysis for obtaining relevant solutions.
\\
\indent We first look at asymptotic behaviors of scalar fields near all of the $AdS_5$ critical points. These behaviors give information about possible types of deformations for the SCFT dual to each $AdS_5$ critical point. It is convenient to redefine the coupling constants as in \eqref{g_redef} together with
\begin{equation}
h_2=\frac{g}{\zeta}
\end{equation}
for $0<\zeta<1$ in order to have $h_2>h_1>0$. Linearizing the BPS equations leads to the following results:
\\
\textbf{Critical point I}\\
\indent Near critical point I, the BPS equations give, for $\rho\neq 2$,
\begin{equation}
\Sigma\sim \phi_1\sim\phi_2\sim e^{-gr}=e^{-\frac{2r}{L_{\textrm{I}}}},\qquad \phi_3\sim e^{-g\left(1-\frac{1}{\rho}\right)r}=e^{-2\left(1-\frac{1}{\rho}\right)\frac{r}{L_{\textrm{I}}}}
\end{equation}
with $L_{\textrm{I}}=\frac{2}{g}$. We see that $\Sigma$, $\phi_1$ and $\phi_2$ are dual to operators of dimension $\Delta=2$ while $\phi_3$ is dual to an operator of dimension $\Delta=2+\frac{2}{\rho}$.
\\
\indent As pointed out in \cite{5D_N4_flow_Davide}, $\Sigma$, $\phi_1$ and $\phi_2$ correspond to the vev of relevant operators with dimension $2$ while $\phi_3$ leads to a source term of an operator with dimension $2+\frac{2}{\rho}$ in the dual $N=2$ SCFT. For $\rho=2$, however, the cubic term of the form $\Sigma \phi_3^2$ of the  expansion of the superpotential is of the same order as the quadratic term $\Sigma^2$ as pointed out in \cite{FGPP}. In this case, the asymptotic behaviors are given by
\begin{eqnarray}
\phi_1\sim\phi_2\sim e^{-\frac{2r}{L_{\textrm{I}}}},\qquad \phi_3\sim C_3e^{-\frac{r}{L_{\textrm{I}}}},\qquad\Sigma\sim Ce^{-\frac{2r}{L_{\textrm{I}}}}+\frac{4C_3^2}{3L_{\textrm{I}}}re^{-\frac{2r}{L_{\textrm{I}}}}
\end{eqnarray}  
which indicates that $\Sigma$ and $\phi_3$ corresponds to source terms of dimension-$2$ and -$3$ operators as also shown in \cite{5D_N4_flow_Davide}.  
\\
\textbf{Critical point II}\\
\indent For $\rho\neq 2$, we find
\begin{eqnarray}
& &\Sigma\sim e^{-\frac{gr}{(1-\zeta)^{\frac{1}{3}}}}=e^{-\frac{2r}{L_{\textrm{II}}}},\qquad \phi_1\sim\phi_2\sim e^{\frac{gr}{(1-\zeta^2)^{\frac{1}{3}}}}= e^{\frac{2r}{L_{\textrm{II}}}},\nonumber \\
& &\phi_3\sim e^{-\frac{gr(\rho-1)}{(1-\zeta^2)^{\frac{1}{3}}}}=e^{-2\left(1-\frac{1}{\rho}\right)\frac{r}{L_{\textrm{II}}}}
\end{eqnarray}
with $L_{\textrm{II}}=\frac{2(1-\zeta^2)^{\frac{1}{3}}}{g}$. At this critical point, $\Sigma$ and $\phi_3$ are still dual respectively to operators of dimensions $\Delta=2$ and $\Delta=2+\frac{2}{\rho}$ as in the previous case. On the other hand, $\phi_1$ and $\phi_2$ are now dual to irrelevant operators of dimension $\Delta=6$.  
\\
\indent As in critical point I, these behavors suggest that $\Sigma$ and $\phi_3$ correspond respectively to a vev and a source term of operators with dimensions $2$ and $2+\frac{2}{\rho}$. For $\rho=2$, the asymptotic expansion gives
\begin{eqnarray}
\phi_1\sim\phi_2\sim e^{\frac{2r}{L_{\textrm{II}}}},\qquad \phi_3\sim C_3e^{-\frac{r}{L_{\textrm{II}}}},\qquad \Sigma\sim Ce^{-\frac{2r}{L_{\textrm{II}}}}+\frac{4C_3^2}{3L_{\textrm{II}}(1-\zeta^2)^{\frac{1}{6}}}re^{-\frac{2r}{L_{\textrm{II}}}}
 \end{eqnarray}
which again implies source terms for operators of dimension $2$ and $3$ dual to $\Sigma$ and $\phi_3$, respectively.  
\\
\textbf{Critical point III}\\
\indent Near this critical point, we have
\begin{eqnarray}
& &\phi_1\sim e^{-\frac{gr(2+\rho)}{3\rho^{\frac{1}{3}}}}=e^{-\frac{2r}{L_{\textrm{III}}}},\qquad \phi_2\sim e^{-\frac{gr}{\rho^{\frac{1}{3}}}}= e^{-\frac{6}{2+\rho}\frac{r}{L_{\textrm{III}}}},\nonumber \\
& &\Sigma\sim\phi_3\sim C_1 e^{-\left(1+\sqrt{\frac{25\rho-22}{2+\rho}}\right)\frac{r}{L_{\textrm{III}}}}+C_2e^{-\left(1-\sqrt{\frac{25\rho-22}{2+\rho}}\right)\frac{r}{L_{\textrm{III}}}}
\end{eqnarray}
with $L_{\textrm{III}}=\frac{6\rho^{\frac{1}{3}}}{g(2+\rho)}$ and constants $C_1$ and $C_2$. In this case, $\phi_1$ and $\phi_2$ are dual to operators of dimensions $\Delta=2$ and $\Delta=\frac{6}{2+\rho}, \frac{2+4\rho}{2+\rho}$, respectively. On the other hand, $\Sigma$ and $\phi_3$ are dual to combinations of operators of dimensions $\Delta=1+\sqrt{\frac{25\rho-22}{2+\rho}}$ and $\Delta=3+\sqrt{\frac{25\rho-22}{2+\rho}}$ as also pointed out in \cite{5D_N4_flow_Davide}.
\\
\textbf{Critical point IV}\\
\indent Finally, we find the behaviors near critical point IV
\begin{eqnarray}
& &\phi_1, \phi_2\sim C_1e^{-\left(1+\sqrt{\frac{\rho+26}{2+\rho}}\right)\frac{r}{L_{\textrm{IV}}}}+C_2 e^{-\left(1-\sqrt{\frac{\rho+26}{2+\rho}}\right)\frac{r}{L_{\textrm{IV}}}},\nonumber \\
& &\Sigma,\phi_3\sim C_3e^{-\left(1+\sqrt{\frac{25\rho-22}{2+\rho}}\right)\frac{r}{L_{\textrm{IV}}}}+C_4 e^{-\left(1-\sqrt{\frac{25\rho-22}{2+\rho}}\right)\frac{r}{L_{\textrm{IV}}}}
\end{eqnarray}
with $L_{\textrm{IV}}=\frac{6\rho^{\frac{1}{3}}(1-\zeta^2)^{\frac{1}{3}}}{g(2+\rho)}$. $\Sigma$ and $\phi_3$ are still dual to combinations of operators of dimensions $\Delta=1+\sqrt{\frac{25\rho-22}{2+\rho}}$ and $\Delta=3+\sqrt{\frac{25\rho-22}{2+\rho}}$ as in the case of critical point III while $\phi_1$ and $\phi_2$ are dual to combinations of operators of dimensions $\Delta=1+\sqrt{\frac{\rho+26}{2+\rho}}$ and $\Delta=3+\sqrt{\frac{\rho+26}{2+\rho}}$.
\\
\indent We note that these behaviors give conformal dimensions consistent with the scalar masses given in the previous section. We are now in a position to give numerical RG flow solutions to the BPS equations. First of all, we take the numerical values of the coupling constants to be 
\begin{equation} 
g=1,\qquad \rho=3,\qquad \zeta=\frac{1}{2}\, .
\end{equation}
There exists a family of RG flows from the $N=4$ critical point with $SO(2)_D\times SO(3)\times SO(3)$ symmetry (critical point I) to the $N=2$ fixed point with $SO(2)_{\textrm{diag}}$ symmetry (critical point IV) as shown in figure \ref{fig1}. In this family, there is an RG flow that begins at critical point I and proceeds directly to critical point IV as shown by the green line in figure \ref{fig1}. In addition, there is a number of RG flows that pass arbitrarily close to critical point II before ending at critical point IV shown by the red, blue and cyan lines in the figure. For clarity, we have also included the dashed lines referring to the values of the corresponding fields at each critical point. These RG flows constitute a family or a cone of flows, in the terminology of \cite{Warner_4D_flow}, bounded by the direct flow (green line) and the limiting flow that passes very close to critical point II and ends at critical point IV (cyan line). All of these RG flows are driven by operators of dimensions $2$ and $2+\frac{2}{\rho}$ in the $N=4$ UV conformal fixed point to $N=2$ critical point IV in the IR.     

\begin{figure}
         \centering
               \begin{subfigure}[b]{0.35\textwidth}
                 \includegraphics[width=\textwidth]{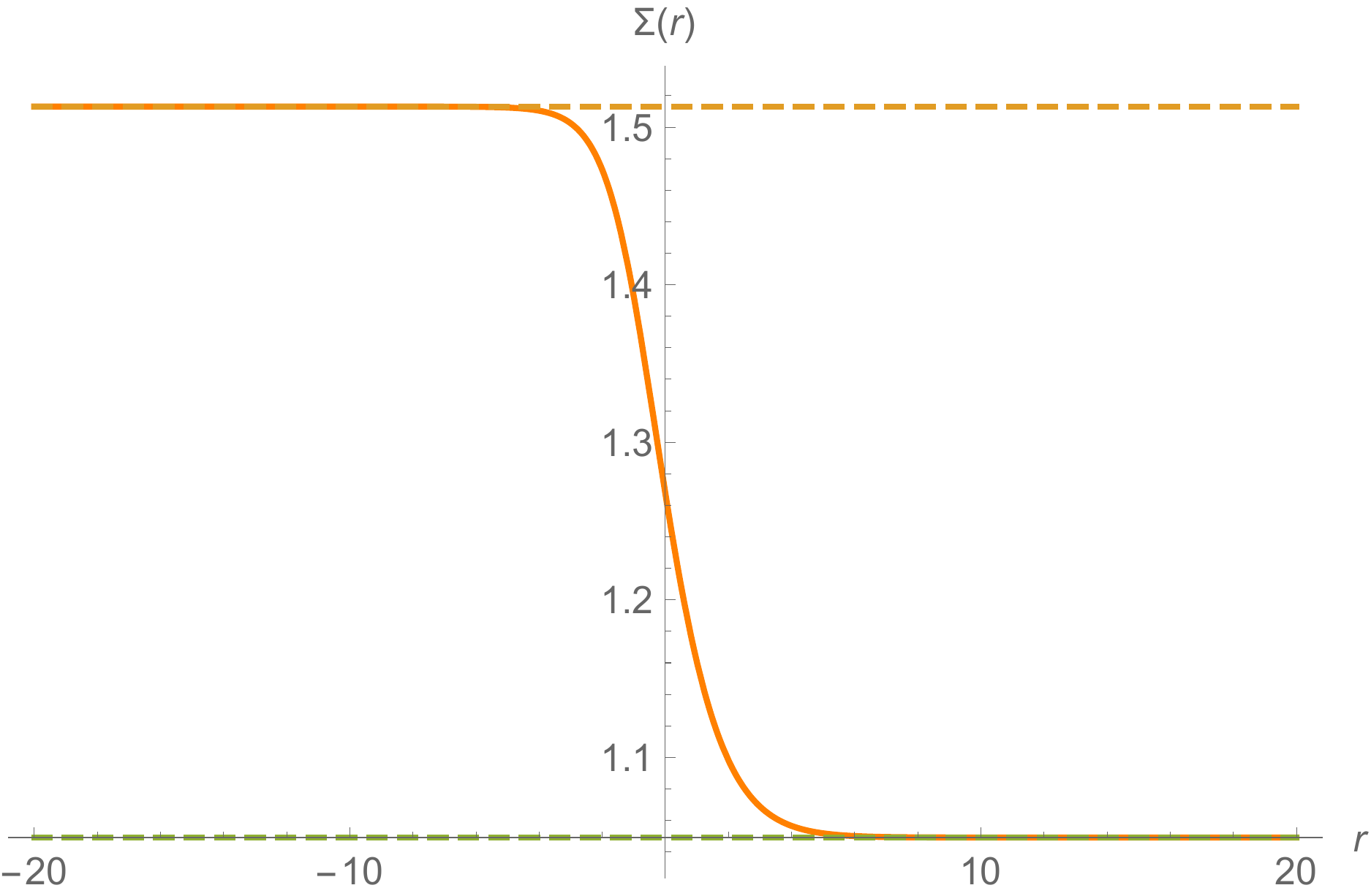}
                 \caption{Solution for $\Sigma(r)$}
         \end{subfigure}
         \begin{subfigure}[b]{0.35\textwidth}
                 \includegraphics[width=\textwidth]{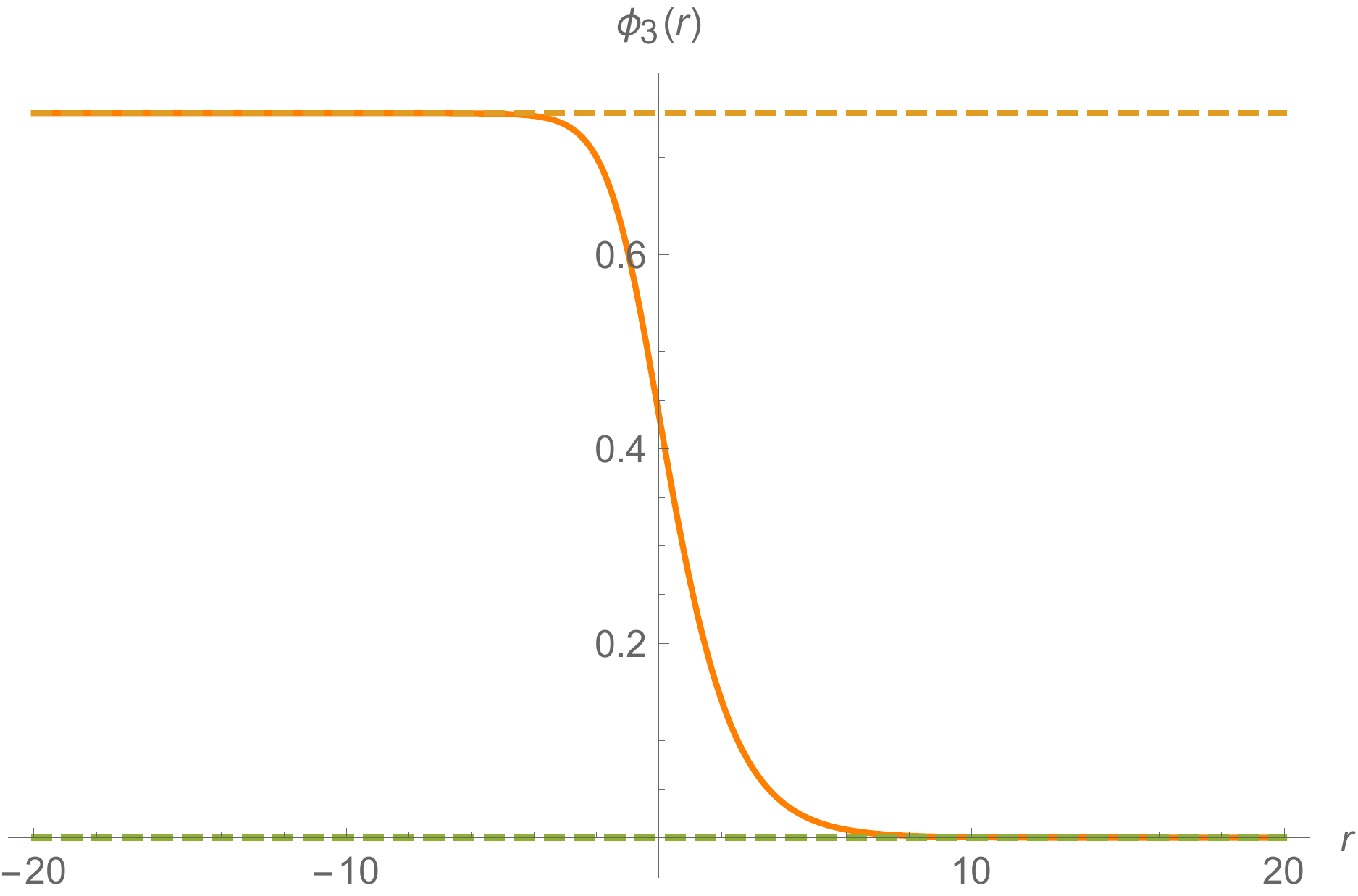}
                 \caption{Solution for $\phi_3(r)$}
         \end{subfigure}
          \begin{subfigure}[b]{0.35\textwidth}
                 \includegraphics[width=\textwidth]{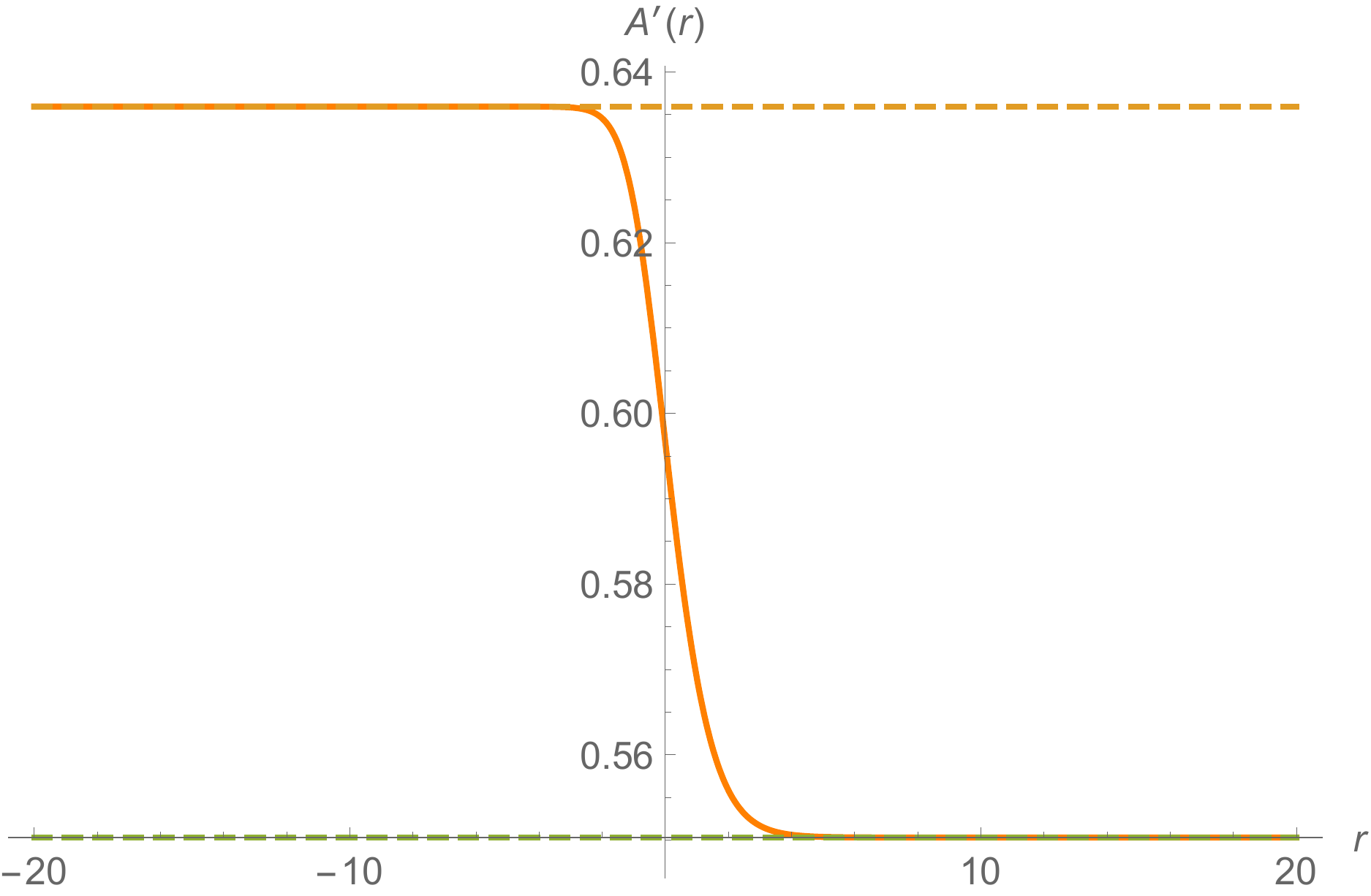}
                 \caption{Solution for $A'(r)$}
         \end{subfigure}
         \caption{An RG flow from $N=4$ $AdS_5$ critical point with $SO(2)_D\times SO(3)_{\textrm{diag}}$ symmetry to $N=2$ $AdS_5$ fixed point with $SO(2)_{\textrm{diag}}$ symmetry in the IR with $g=1$, $\rho=3$ and $\zeta=\frac{1}{2}$. In this solution, $\phi_1=\phi_2=\frac{1}{2}\ln\left[\frac{1-\zeta}{1+\zeta}\right]$ along the flow}\label{fig2}
 \end{figure}

\begin{figure}
         \centering
               \begin{subfigure}[b]{0.35\textwidth}
                 \includegraphics[width=\textwidth]{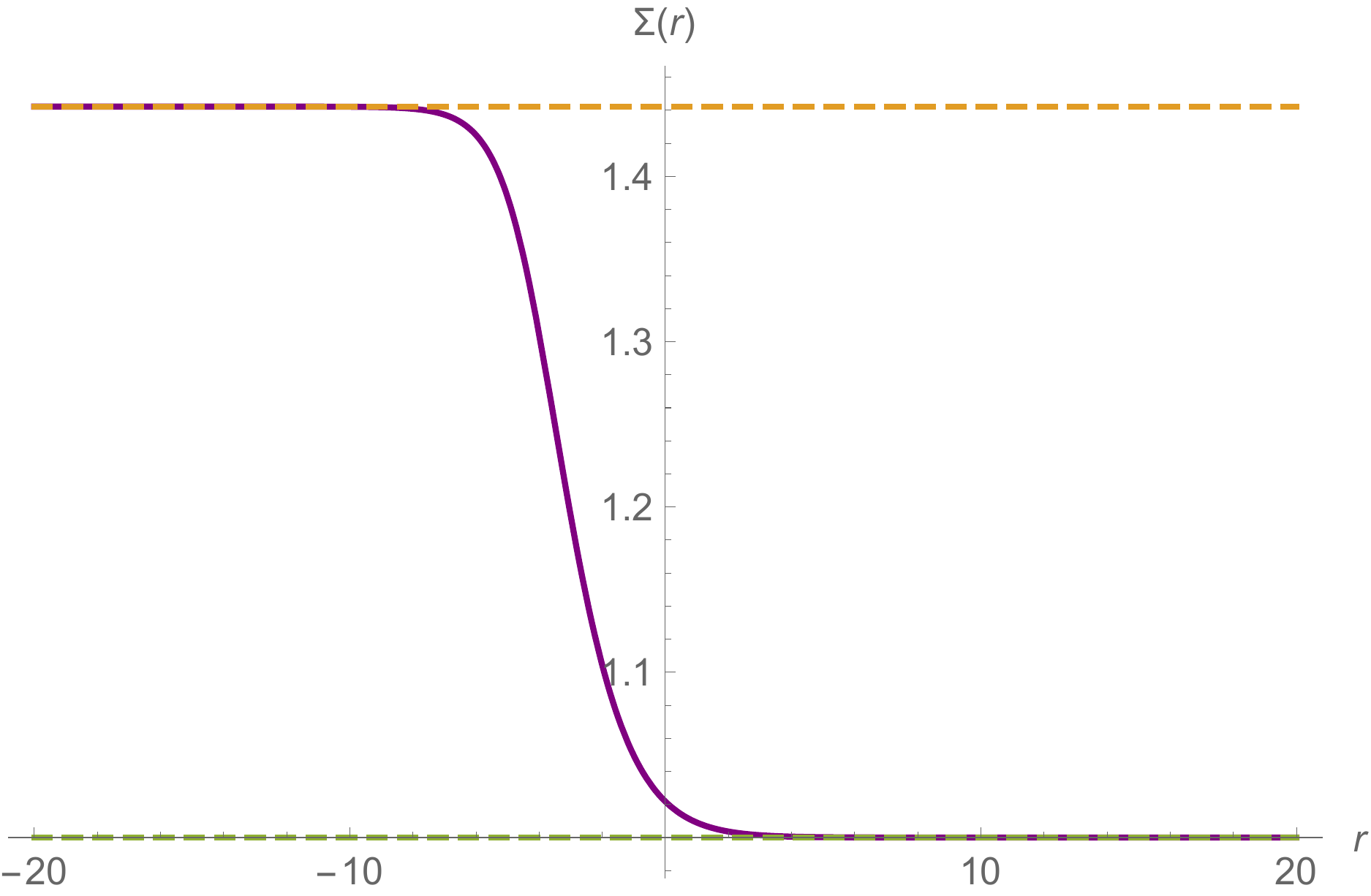}
                 \caption{Solution for $\Sigma(r)$}
         \end{subfigure}
         \begin{subfigure}[b]{0.35\textwidth}
                 \includegraphics[width=\textwidth]{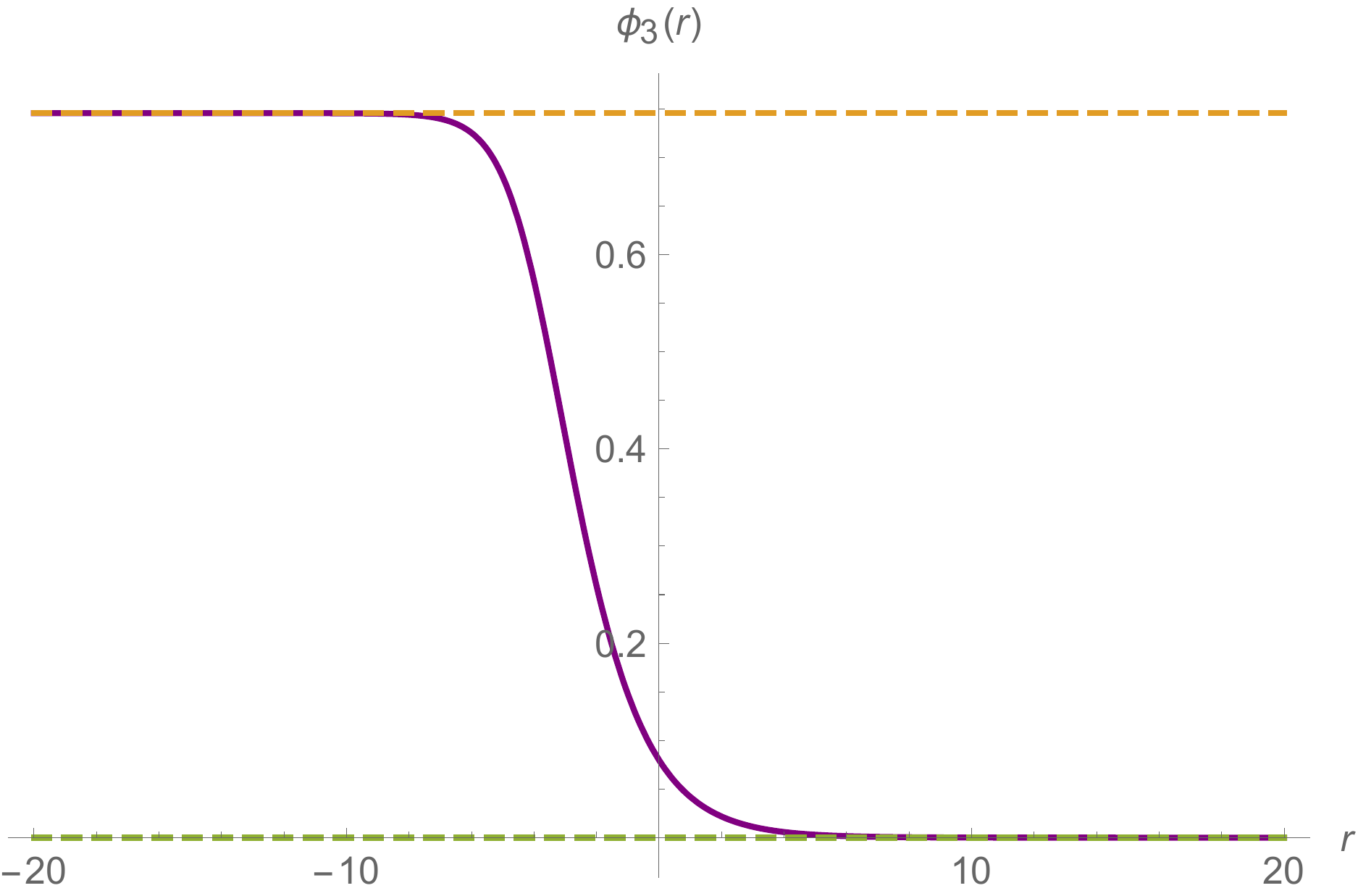}
                 \caption{Solution for $\phi_3(r)$}
         \end{subfigure}
          \begin{subfigure}[b]{0.35\textwidth}
                 \includegraphics[width=\textwidth]{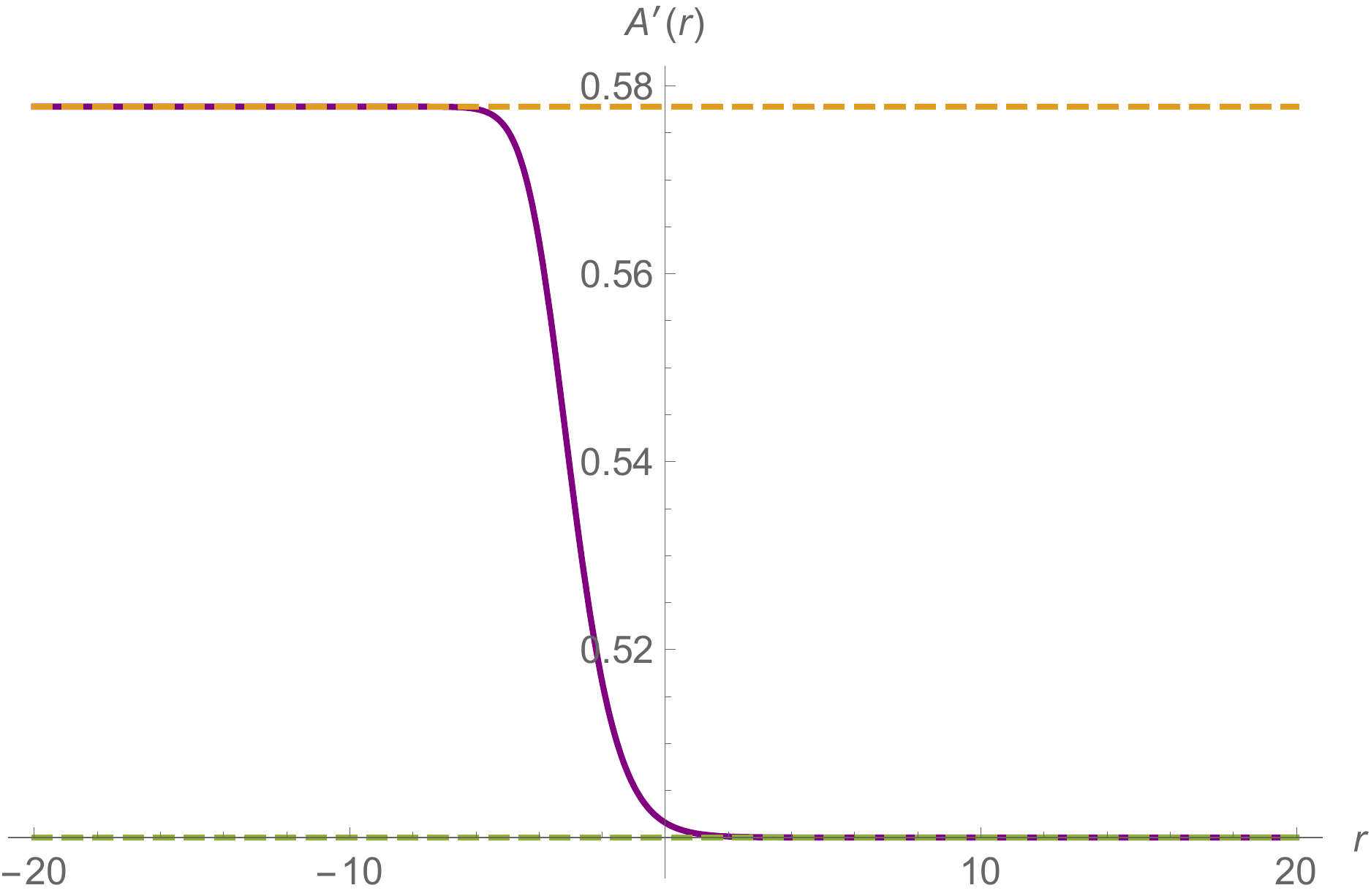}
                 \caption{Solution for $A'(r)$}
         \end{subfigure}
         \caption{An RG flow from $N=4$ $AdS_5$ critical point with $SO(2)_D\times SO(3)\times SO(3)$ symmetry to $N=2$ $AdS_5$ fixed point with $SO(2)_{\textrm{diag}}\times SO(3)$ symmetry in the IR with $g=1$, $\rho=3$ and $\zeta=\frac{1}{2}$. In this case, $\phi_1=\phi_2=0$ along the flow}\label{fig3}
 \end{figure}

There is also a direct RG flow from critical point II to critical point IV as shown in figure \ref{fig2}. We have consistently set $\phi_1=\phi_2=\frac{1}{2}\ln\left[\frac{1-\zeta}{1+\zeta}\right]$ along the flow. Although this simplifies the BPS equations considerably, we are not able to find an analytic flow solution. As in figure \ref{fig1}, the dashed lines refer to the values at the critical points. This RG flow is driven by relevant operators of dimensions $2$ and $2+\frac{2}{\rho}$ dual respectively to $\Sigma$ and $\phi_3$. The solution is similar to the $N=2$ RG flow from $N=4$ critical point I to $N=2$ critical point III given in \cite{5D_N4_flow_Davide} with $\phi_1=\phi_2=0$ along the flow. We also show this solution in figure \ref{fig3} for the sake of comparison.      

 As a final note, the flow in figure \ref{fig2} could joint the $N=4$ RG flow from critical point I to critical point II to form a cascade of RG flows. On this flow, the operators dual to $\phi_1$ and $\phi_2$ become irrelevant at an intermediate $N=4$ fixed point (critical point II). The relevant operators ($\Delta=2$ and $\Delta=2+\frac{2}{\rho}$) dual to $\Sigma$ and $\phi_3$ then further drive the flow to critical point IV. The resulting solution is similar to the cyan line in figure \ref{fig1}.  

\section{Conclusions and discussions}\label{conclusion}
In this paper, we have studied five-dimensional $N=4$ gauged supergravity with $SO(2)_D\times SO(3)\times SO(3)$ gauge group. Within the $SO(2)_{\textrm{diag}}$ invariant scalar sector, we have identified four supersymmetric $AdS_5$ vacua with $N=4$ and $N=2$ supersymmetries. One of the $N=4$ vacua is the trivial critical point located at the origin of the scalar manifold and preserves the full $SO(2)_D\times SO(3)\times SO(3)$ gauge symmetry. The other $N=4$ vacuum only preserves $SO(2)_D\times SO(3)_{\textrm{diag}}$ symmetry. Both of these vacua are dual to $N=2$ SCFTs in four dimensions with $SO(3)$ and trivial flavor symmetries, respectively. The remaining two $AdS_5$ vacua are $N=2$ supersymmetric and are dual to $N=1$ SCFTs in four dimensions. One of these critical points has $SO(2)_{\textrm{diag}}\times SO(3)$ symmetry. All of these three $AdS_5$ vacua have an analogue in related gauge groups $SO(2)\times SO(3)\times SO(3)$ and $SO(2)_D\times SO(3)$ studied in \cite{5D_N4_flow_Davide} and \cite{5D_N4_flow}.    
\\
\indent In addition, we have found a genuinely new $N=2$ $AdS_5$ vacuum with $SO(2)_{\textrm{diag}}$ symmetry. We have also computed the full scalar masses at all of the aforementioned $AdS_5$ vacua and studied holographic RG flows interpolating among these critical points. Apart from the known $N=4$ RG flows between $N=4$ critical points and the $N=2$ RG flow from the trivial $N=4$ critical point to $SO(2)_{\textrm{diag}}\times SO(3)$ $N=2$ critical point, we have found a family of RG flows between the trivial $N=4$ critical point to the new $SO(2)_{\textrm{diag}}$ $N=2$ critical point. Some of these RG flows pass arbitrarily close to the $SO(2)_D\times SO(3)_{\textrm{diag}}$ $N=4$ critical point in addition to the direct RG flow from the trivial $N=4$ critical point to the $SO(2)_{\textrm{diag}}$ $N=2$ critical point. We have numerically given all of these RG flows. The results could give a holographic description of the possible RG flows between various conformal phases of strongly coupled $N=1$ and $N=2$ SCFTs in four dimensions and might be useful in other related study.
\\
\indent It would be interesting to identify precisely the dual $N=1$ and $N=2$ SCFTs dual to all the above $AdS_5$ vacua and the corresponding RG flows. Since the $SO(2)_D\times SO(3)\times SO(3)$ gauged supergravity under consideration here has currently no known higher-dimensional origin, it is of particular interest to find the embedding of this $N=4$ gauged supergravity in string/M-theory in which a complete holographic description can be obtained. This might be achieved by using recent results in exceptional field theories and double field theories. In particular, in \cite{Malek_AdS5_N4_embed}, consistent truncations of $N=4$ gauged supergravity with $n\leq 3$ vector multiplets from eleven-dimensional supergravity have been shown. In this case, the $N=4$ $AdS_5$ vacua are embeddable in the maximal $SO(5)$ gauged supergravity in seven dimensions. The extension of this result to include $AdS_5$ vacua that cannot be embedded in the seven-dimensional theory might possibly lead to the embedding of the $N=4$ gauged supergravity with $SO(2)_D\times SO(3)\times SO(3)$ gauge group. Finally, other types of holographic solutions such as black strings, black holes and Janus solutions within the $N=4$ gauged supergravity studied here are also worth consideration. This could be done along the same line as in the recent results \cite{5D_N4_flow}, \cite{5D_N4_BH} and \cite{Bobev_5D_Janus2}.                 
\vspace{0.5cm}\\
{\large{\textbf{Acknowledgement}}} \\
This work is funded by National Research Council of Thailand (NRCT) and Chulalongkorn University under grant N42A650263.

\end{document}